\documentclass[showpacs,preprintnumbers,amsmath,amssymb,floatfix]{revtex4}
\usepackage[german, english]{babel}
\usepackage{graphicx}
\usepackage{amsmath}
\usepackage{amsfonts}
\usepackage{amssymb}
\usepackage{epsfig}
\usepackage[hang,nooneline]{subfigure}
\newcommand{\insertplot}[5]{\begin{figure}
 \hfill\hbox to 0.05in{\vbox to #5in{\vfill
 \inputplot{#1}{#4}{#5}}\hfill}
 \hfill\vspace{-.1in}
 \caption{#2}\label{#3}
 \end{figure}}
\newcommand{\inputplot}[3]{
 \special{ps: plotfile #1}
}
 
\newcounter{fig}   

\newcommand{\vphi}{\varphi}

\begin{document}

\title{\bf Rotating Wormholes in Five Dimensions}
\author{{\bf Vladimir Dzhunushaliev$^{1,2,3}$}}
\email[{\it Email:}]{v.dzhunushaliev@gmail.com}
\author{{\bf Vladimir Folomeev$^{1,3}$}}
\email[{\it Email:}]{vfolomeev@mail.ru}
\author{{\bf  Burkhard Kleihaus $^1$}}
\email[{\it Email:}]{b.kleihaus@uni-oldenburg.de}
\author{{\bf Jutta Kunz$1^1$}}
\email[{\it Email:}]{jutta.kunz@uni-oldenburg.de}
\author{{\bf Eugen Radu$^1$}}
\email[{\it Email:}]{eugen.radu@uni-oldenburg.de}
\affiliation{$^1$
Institut f\"ur Physik, Universit\"at Oldenburg, Postfach 2503,
D-26111 Oldenburg, Germany\\
$^2$
Dept. Theor. and Nucl. Phys., KazNU, Almaty, 010008, Kazakhstan\\
$^3$Institute of Physicotechnical Problems and Material Science of the NAS
of the
Kyrgyz Republic, 265 a, Chui Street, Bishkek, 720071,  Kyrgyz Republic 
}

\date{\today}

\begin{abstract}
We consider rotating Lorentzian wormholes with a phantom field in five dimensions.
These wormhole solutions 
possess equal
angular momenta and thus represent cohomogeneity-1  configurations.
For a given size of the throat, the angular momenta are bounded
by the value of the corresponding extremal Myers-Perry black hole,
which represents the limiting configuration.
With increasing angular momenta the throat becomes increasingly deformed.
At the same time, the violation of the null energy condition
decreases to zero, as the limiting configuration is approached.
Symmetric wormhole solutions satisfy a Smarr-like relation, which is analogous
to the Smarr relation of extremal black holes.
A stability analysis shows that the unstable mode of the static 
wormholes solutions vanishes when the angular momentum exceeds 
some critical value. 
\end{abstract}

\maketitle

\section{Introduction}

Discovered long ago, an `Einstein-Rosen bridge' \cite{Einstein:1935tc} represents a connection
between the two exterior regions of a 
Schwarzschild
black hole space time, and thus forms an inter-universe connection.
As realized by
Wheeler \cite{Wheeler:1957mu,Wheeler:1962} 
such a bridge could also form an intra-universe connection,
bridging 
 arbitrarily large distances 
within a single universe.
However, 
it became soon clear
that such a Schwarzschild wormhole 
would not be traversable for matter and for light 
\cite{Kruskal:1959vx,Fuller:1962zza,Eardley:1974zz,Redmount:1985,Wald:1980nk}.

A different type of wormhole 
that is not associated with black holes
was discovered by Ellis \cite{Ellis:1973yv,Ellis:1979bh}, Bronnikov \cite{Bronnikov:1973fh},
and Kodama \cite{Kodama:1978dw}.
To obtain such wormhole solutions, they had to invoke
some form of exotic matter,
whose energy-momentum tensor violates the null,
weak and strong energy conditions.
The candidate they chose for such exotic matter 
was a phantom scalar field, whose kinetic term has a reversed sign.

In the seminal work of Morris and Thorne \cite{Morris:1988cz} 
traversability of this type of wormhole was the central goal.
However, traversability does not only require
small accelerations and tidal forces,
but it  also requires stability of the wormhole.
Whereas early work seemed to indicate
that such phantom field wormholes could be stable
\cite{Kodama:1978dw,ArmendarizPicon:2002km},
later work showed, that they are unstable
\cite{Shinkai:2002gv,Gonzalez:2008wd,Gonzalez:2008xk,Bronnikov:2011if,Dzhunushaliev:2013lna,Charalampidis:2013ixa}.

To achieve stability, one may on the one hand consider a 
generalized theory of gravity with a different type of matter content.
A case at hand is the study of wormholes in Einstein-Gau\ss -Bonnet-dilaton
gravity \cite{Kanti:2011jz,Kanti:2011yv}.
On the other hand, it has been argued, that rotating wormholes would
have a higher possibility of being stable \cite{Matos:2005uh}
and thus traversable.

A first example of a rotating 
wormhole was given by Teo \cite{Teo:1998dp}.
However,
he did not solve a set of Einstein-matter equations,
but discussed only the general form and properties of such a wormhole.
At the same time
Khatsymovsky \cite{Khatsymovsky:1998tv} discussed 
some general properties of
a slowly rotating wormhole.
A thorough analysis of the Teo wormhole
was given in \cite{PerezBergliaffa:2000ms} and \cite{Kuhfittig:2003wr}.

Recently, Kashargin and Sushkov \cite{Kashargin:2007mm,Kashargin:2008pk}
set out to construct self-consistent wormhole solutions.
Invoking a phantom scalar field
they solved the Einstein-matter equations
for slow rotation in first and second order. 
Thus these perturbative solutions represent slowly rotating 
generalizations of the Morris-Thorne wormholes.
They showed that the mass of a rotating wormhole
is greater than that of a nonrotating one,
and that the null energy condition violation is weaker in a
rotating wormhole.

Clearly, the outstanding quest is to obtain  nonperturbative
rotating wormhole solutions and to subsequently study their stability.
However, in four dimensions one has to deal with partial differential equations
in the presence of rotation. 
Therefore we here consider  the simpler problem
of rotating wormholes in five dimensions,
whose two angular momenta are of equal magnitude.
In this case
one obtains a cohomogeneity-1 problem, and thus 
only a system of ordinary differential equations 
\cite{Kunz:2005nm,Hartmann:2010pm}.

While the non-rotating wormhole solutions
with a phantom scalar field in five dimensions
can be given in analytical form \cite{Shinkai:2013}, we obtain 
only partially analytical solutions in the rotating case.
Here two of the metric functions are obtained numerically.
We find a Smarr-type relation for the
rotating wormholes, that has the same form as 
the Smarr relation for extremal rotating black holes.
We determine the domain of existence of the rotating wormholes
and the change of the geometry of the throat with 
increasing angular momentum.
Subsequently we analyze the
stability of these rotating wormhole solutions.

The paper is organized as follows.
In section II, we present
the action, the Ans\"atze and the field equations.
We discuss the main wormhole features in section III.
The analytical solutions for the
static wormholes and the numerical results for the 
rotating wormholes
are presented in section IV.
Section V is devoted to the stability analysis of these solutions.
We give our conclusions in section VI.
The Appendix A contains a generalization of the analytical solution  for the
static wormholes for an arbitrary dimension $D\geq 4$.
In Appendix B we give the asymptotics of the solutions together
with their expression in the slowly rotating limit. 
The Appendix C contains
 an exact solution describing
a spinning wormhole interpolating between a 
Kaluza-Klein monopole background and a squashed
 $AdS_2\times S^3$ spacetime.

\section{Action and Field Equations}
\subsection{Action}
We consider Einstein gravity coupled to a phantom field in five dimensions. 
The action 
\begin{equation}
S=\int \left[ \frac{1}{16\pi G}{\cal R} + 
{\cal L}_{\rm ph}  \right]\sqrt{-g}d^5x  
 \label{action}
\end{equation}
consists of the Einstein-Hilbert action
with curvature scalar $\cal R$, five-dimensional gravitational constant $G$ and determinant of the 
metric $g$, and the Lagrangian of the phantom field $\phi$  
\begin{equation}
 {\cal L}_{\rm ph} = \frac{1}{2}\partial_\mu \phi\partial^\mu \phi \ .
\label{lphi}
\end{equation}

Variation of the action with respect to the metric
leads to the Einstein equations
\begin{equation}
G_{\mu\nu}= {\cal R}_{\mu\nu}-\frac{1}{2}g_{\mu\nu}{\cal R} = 8\pi G T_{\mu\nu}
\label{ee} 
\end{equation}
with stress-energy tensor
\begin{equation}
T_{\mu\nu} = g_{\mu\nu}{{\cal L}}_{\rm ph}
-2 \frac{\partial {{\cal L}}_{\rm ph}}{\partial g^{\mu\nu}} \ ,
\label{tmunu} 
\end{equation}
%
\subsection{Ans\"atze}
The static, spherically symmetric wormhole solutions  
can be studied by using  a metric ansatz
\begin{equation}
\label{metric-static}
ds^2=-U_0(l) dt^2 +U_1(l)\left(dl^2+h(l)d\Omega_3^2 \right) 
\end{equation}
with
\begin{equation}
h(l)=l^2+r_0^2~,
\end{equation}
where
the coordinate $l$ takes positive and negative 
values, $-\infty< l < \infty$, while $t$ is the time coordinate. 
The wormhole throat is located at $l=0$,
where $U_0>0$.
 The limits $l\to \pm\infty$
correspond to two disjoint asymptotic regions.
Also, $d\Omega_3^2$ is the metric of the round three-sphere, 
$d\Omega_3^2=\frac{1}{4}(\sigma_1^2+\sigma_2^2+\sigma_3^2)$,
with the left-invariant 
1-forms $\sigma_i$ on $S^3$, 
$\sigma_1=\cos \bar\psi d\bar \theta+\sin\bar\psi \sin \bar \theta d \bar\varphi$,
$\sigma_2=-\sin \bar \psi d\bar \theta+\cos\bar\psi \sin \bar \theta d\bar \varphi$,
$\sigma_3=d\bar\psi  + \cos \bar \theta d \bar\varphi$  
 and $(\bar \theta,\bar \varphi, \bar \psi)$ the Euler angles.

A simple rotating generalization of the metric ansatz (\ref{metric-static})
which leads to cohomogeneity-1  configurations is found by taking
\begin{eqnarray}
\label{rot1}
ds^2=-U_0(l)  dt^2+U_1(l) \left(dl^2+ h(l)\frac{1}{4}(\sigma_1^2+\sigma_2^2) \right)
+U_2(l)h(l)\frac{1}{4}(\sigma_3-2 \omega(l) dt)^2.
\end{eqnarray}
A more usual form of this metric ansatz
is found by defining $\bar \theta =2\theta$,
$\bar \varphi=\varphi-\psi$, $\bar \phi=\varphi+\psi$, with 
$0\leq \theta\leq \pi/2$, 
$0\leq \varphi< 2\pi $, 
$0\leq \psi< 2\pi $.
Also, it is convenient to take
$U_0(l)=e^{2a(l)}$, 
$U_1(l)=p(l) e^{-q(l)}$,
$U_2(l)=p(l) e^{q(l)-2a(l)}$
which leads to partially analytical solutions.

With this choice, the metric ansatz we employ
for the study of stationary rotating  wormhole solutions with two equal
angular momenta  reads
\begin{eqnarray}
ds^2 & = &  -e^{2a} dt^2 + p e^{-a}\left\{
e^{a-q}\left[dl^2 +h d\theta^2\right]
                   + e^{q-a}
		   h\left[ \sin^2\theta (d\varphi -\omega dt)^2
		         +\cos^2\theta (d\psi -\omega dt)^2\right]\right.
\nonumber\\
& &			\left.
			+\left(e^{a-q}- e^{q-a}\right)
			h \sin^2\theta\cos^2\theta(d\psi -d\varphi)^2
			\right\} \ .
\label{lineel}
\end{eqnarray}
%
The scalar phantom field $\phi$ is also a function of $l$ only,
\begin{eqnarray}
\phi=\phi(l).
\end{eqnarray}

\subsection{Einstein and Phantom Field Equations}
For the above Ansatz the phantom field equation reduces to
\begin{equation}
\nabla_\mu \nabla^\mu \phi = 0 \ \
\Longleftrightarrow  \ \
\left(p\sqrt{h^3}\phi'\right)' = 0 \ ,
\label{phieq}
\end{equation}
 where a prime denotes $d/d l$.
Consequently, 
\begin{equation}
\phi' = \frac{Q}{p\sqrt{h^3}} \ ,
\label{phiprime}
\end{equation}
where $Q$ is a constant.

The Einstein equations can be written in the form 
$R_{\mu\nu} = - 8\pi G \partial_\mu \phi\partial_\nu \phi$, which reduces 
to 
\begin{eqnarray}
R_{ll} & = & - 8\pi G \frac{Q^2}{ h^3 p^2} \ ,
\label{ric22} \\
R_{\mu\nu} & = & 0 \ \ \ \ {\rm for} \ \mu\neq l\ , \ \nu\neq l \ ,
\label{Eineqs}
\end{eqnarray}
  with 
\begin{eqnarray}
\label{Rll-expr}
R_{ll} & = & -\frac{3p''}{2p}
+\frac{q''}{2}
-\frac{3h''}{2h}
+\frac{3h'^2}{4h^2}
+\frac{3p'^2}{2p^2}
-2a'^2
-\frac{ q'^2}{2}
+a'q' 
\nonumber\\
& &
 -\frac{p'}{p}(\frac{q'}{2}-a')
-\frac{h'}{4h}(q'+ \frac{3p'}{p}-4a' )
+\frac{1}{2}e^{q-4a}\,h\,p\, \omega'^2 \ .
\end{eqnarray}
In Eq.~(\ref{ric22}) we substituted $\phi'$ from Eq.~(\ref{phiprime}),
thus we can express
$Q^2$ in terms of the metric functions.
We will treat Eq.~(\ref{ric22}) as a constraint.

We note that the equation
\begin{equation}
\sqrt{-g}R^t_\vphi = 0 \ \
\Longleftrightarrow  \ \
 \left(\sqrt{h^5}p^2 e^{q-4a}\omega'\right)' = 0 
\label{omeq}
\end{equation}
has the solution
\begin{equation}
\omega' = \frac{c_\omega}{h^{\frac{5}{2}}p^2 e^{q-4a}} \ ,
\label{omprime}
\end{equation}
where $c_\omega$ is  an integration constant. 

The equation $R^\theta_\theta=0$ yields a remarkably simple
ODE for the function $p$,
\begin{equation}
p'' + \frac{5 l}{l^2+r_0^2} p'-2 \left(\frac{r_0}{l^2+r_0^2}\right)^2 p = 0.
\label{eqp}
\end{equation}
Its general solution is
\begin{equation}
p(l) = -c_1 \frac{l}{\sqrt{l^2+r_0^2}} 
      + c_2\frac{l^2+r_0^2/2}{l^2+r_0^2} \ ,
\label{solp1}
\end{equation}
where $c_1$ and $c_2$ are constants.
To obtain asymptotically flat solutions in the limit $l \to \pm \infty$,
we set $c_1=0$ and $c_2=1$, which yields
%
\begin{equation}
p(l) = \frac{l^2+r_0^2/2}{l^2+r_0^2} \ .
\label{solp}
\end{equation}
This leads to the scalar field expression 
\begin{equation}
\phi(l) = \frac{2Q }{ r_0^2} 
\left( 
{\arctan}\frac{l}{\sqrt{l^2+r_0^2}}-\frac{\pi}{4}
\right),
\label{scalar}
\end{equation}
where an integration constant has been chosen 
such that $\phi \to 0$
as $l \to \infty$ (note that $\phi \to  -\pi Q/ r_0^2$ as $l \to -\infty$). 

The equations $R^t_t=0$ and $R^\vphi_\vphi+R^\psi_\psi$
lead to the system of ODEs
\begin{eqnarray}
& & 
\frac{1}{p (l^2+r_0^2)^{\frac{3}{2}}}
\left(p (l^2+r_0^2)^{\frac{3}{2}} a'\right)'
 -\frac{c_\omega^2 e^{4a-q}}{2 p^3(l^2+r_0^2)^4}  =0 \ ,
\label{eqa}\\
& & 
\frac{1}{p (l^2+r_0^2)^{\frac{3}{2}}}
\left(p (l^2+r_0^2)^{\frac{3}{2}} q'\right)'
 -\frac{4 \left(e^{2(q-a)}-1\right)}{(l^2+r_0^2)}  =0 \ .
\label{eqq} 
\end{eqnarray}
Unfortunately, we could not construct a closed form expression for $a$ and $q$
except in the slowly rotating case (see, however, Appendix C).
The nonperturbative solutions are constructed numerically.
In this approach, 
the constraint equation (\ref{ric22}) is used to monitor the quality of the
numerical results.

\section{Physical properties}

\subsection{Asymptotic behavior and the Smarr relation}

Let us begin by noting, that in the case of rotating  wormholes
we cannot simply require asymptotic flatness for both universes,
as already realized in \cite{Khatsymovsky:1998tv}.
We therefore now consider the asymptotic region $l \to \infty$,
where we demand that the spacetime be asymptotically flat.

An expansion in the asymptotic region $l \to \infty$ yields
\begin{eqnarray}
e^{2a} & = & 1 -\frac{8 G}{3 \pi} \frac{M}{l^2} + O(l^{-3}) \ , 
\label{asymp1} \\
\omega  & =  & \frac{4 G}{\pi} \frac{J}{l^4} + O(l^{-5}) \ ,  
\label{asymp2}
\end{eqnarray}
where $M$ and $J=J_1=J_2$ denote the mass and the two equal angular momenta,
respectively, and each angular momentum is associated with rotation in 
an orthogonal plane. From Eq.~(\ref{asymp1}) we find 
\begin{equation}
l^3 a' \to \frac{8 G}{3 \pi} M \ \ \ {\rm as} \ l \to \infty \ .
\label{atoinfty}
\end{equation}
Computing $\omega'$ from Eq.~(\ref{asymp2}) and comparing with 
Eq.~(\ref{omprime}), we can read off the constant $c_\omega$, finding
$c_\omega= -\frac{16 G}{\pi}J$.

To derive the Smarr relation we observe that 
\begin{equation}
\sqrt{-g} R^t_t = 0 \ \
\Longleftrightarrow \ \
\left(2 p (l^2+r_0^2)^{\frac{3}{2}} a' - c_\omega \omega\right)' = 0
\label{rtteq}
\end{equation}
Integration from  the location of the throat $l=l_0$ to $l=\infty$ yields
\begin{equation}
\frac{16 G}{3 \pi} M -
\left(2 \left(p (l^2+r_0^2)^{\frac{3}{2}} a'\right)_0 
     + \frac{16 G}{\pi} J \omega_0\right)
=0 \ , 
\label{Sm1}
\end{equation}
where the subscript $0$ denotes evaluation at $l=l_0$.

For symmetric wormhole solutions 
the throat is located at $l_0=0$,
and the metric function $a$ is symmetric, thus
$a'_0=0$. Eq.~(\ref{Sm1}) then reduces to
\begin{equation}
\frac{16 G}{3 \pi} M =
     \frac{16 G}{\pi} J \omega_0     \ \
\Longleftrightarrow \ \
\frac{2}{3} M =
     2  \omega_0 J \ .
\label{Sm2}
\end{equation}
Note, that the Smarr relation  Eq.~(\ref{Sm2}) for symmetric wormholes agrees with the 
one for  extremal Myers-Perry black holes with equal angular momenta, when the angular velocity $\omega_0$ 
of the throat of the wormhole is replaced
by the horizon angular velocity $\omega_{\rm H}$ of the black hole.

Let us now turn to the discussion of the other asymptotic limit,  $l \to -\infty$.
Clearly, the static solutions are asymptotically flat both in the
symmetric and non-symmetric case. For non-symmetric wormholes,
however, time is flowing at a different rate in the two 
asymptotic regions.
To better understand the asymptotic region $l \to -\infty$
for the rotating solutions,
we first integrate Eq.~(\ref{omprime})
to find the function $\omega(l)$
\begin{equation}
\omega(l)=\omega_0+ c_\omega \int_{l_0}^l \frac{e^{4a-q}}{h^{\frac{5}{2}}p^2 } dl'  \ .
\label{omint}
\end{equation}
Since we require asymptotic flatness in the asymptotic region $l \to \infty$
this means that 
the condition $\omega=0$ must hold for $l \to \infty$. Consequently,
the angular velocity $\omega_0$ 
of the throat of the wormhole is given by
\begin{equation}
\omega_0=- c_\omega \int_{l_0}^\infty  \frac{e^{4a-q}}{h^{\frac{5}{2}}p^2 } dl 
\ .
\label{om0}
\end{equation}
Since $\omega$ is a monotonic function of $l$, we find that
$\omega(-\infty) < \omega_0 < 0$, when $c_\omega$ is positive.
Thus as discussed in \cite{Khatsymovsky:1998tv}
$\omega(-\infty) \ne \omega(\infty)$.
In the case of symmetric wormholes
\begin{equation}
\omega(-\infty)= 2  \omega_0 \ .
\label{om-infty}
\end{equation}
{From} these considerations
we conclude that for wormhole solutions with rotating throat one
of the asymptotic regions has to rotate when the other one is static.
An appropriate coordinate transformation 
for $\phi$ and $\psi$ will make the region $l \to -\infty$
non-rotating, but then the region $l \to \infty$
will be rotating.

\subsection{Geometry of the throat}

We now address the location and the geometry of the throat.
For any fixed $t$ and $l$ the area of the hypersurface is given by 
\begin{equation}
A(l) = \frac{(2\pi)^2}{2} 
     \left[(l^2+r_0^2)^3 p^3 e^{-(2a+q)}\right]^\frac{1}{2} \ .
\label{surface}
\end{equation}
The condition for a minimal area, $A'(l_0)=0$, then determines the location
of the throat. This yields
\begin{equation}
\left(\frac{3 p'}{p} + \frac{6l}{l^2+r_0^2} -2a' -q'\right)_{l=l_0} = 0 \ .
\label{cond_min}
\end{equation}
For solutions symmetric with respect to the transformation
$l \to -l$, the derivatives $a'(0)$, $p'(0)$ and $q'(0)$ vanish.
Consequently,  the throat is located at $l=0$,
and $A(0)$ is the hypersurface of minimal area.

In order to characterize the geometry of the throat we first
consider the embeddings of the planes 
$\theta=0$ and $\theta=\pi/2$.
The metric on such a plane is given by
\begin{equation}
ds^2 = p e^{-q} dl^2 + p(l^2+r_0^2) e^{q-2a}d\vphi^2
= d\rho^2 +dz^2 + \rho^2 d\vphi^2 \ ,
\label{embed1}
\end{equation}
where $\rho$, $z$, and $\vphi$ are cylindrical coordinates on the 
Euclidean embedding space. Regarding $\rho$ and $z$ as functions of 
$l$ we find 
\begin{equation}
\rho(l) = \sqrt{p(l^2+r_0^2) e^{q-2a}} \ ,  \ \ 
z(l) = \int_{l_0}^l{\sqrt{p e^{-q} - \left(\frac{d\rho}{dl}\right)^2}} dl' \ ,
\label{rho-z1}
\end{equation}
which form a parametric representation of the curve $z(\rho)$.
Thus the circumferential radius of the throat in the 
planes of rotation is given by $\rho_c= \rho(l_0)$.

To characterize the deformation of the throat we also
embed a plane which intersects both rotational planes. This can
be achieved by fixing the azimuthal coordinates, 
$\vphi = \vphi_0$, $\vphi = \vphi_0+\pi$ (mod $2\pi$)
and $\psi = \psi_0$, $\psi = \psi_0+\pi$ (mod $2\pi$),\
for some $\vphi_0$ and $\psi_0$.
This yields another circumferential radius of the throat,
$\rho_d= \tilde{\rho}(l_0)$.

The ratio $\rho_c/\rho_d$ then indicates the deformation of the throat
due to rotation.
For wormhole solutions symmetric with
respect to the transformation $l \to -l$, whose throat is located at
$l_0=0$, we find for this ratio $\rho_c/\rho_d = e^{q_0-a_0}$.

\subsection{Energy conditions}

Since violation of the null energy condition (NEC)
implies violation of the weak and strong energy conditions,
we here focus on the NEC, which states that
\begin{equation}
\Xi = T_{\mu\nu} k^\mu k^\nu \ge 0 \ ,
\end{equation}
where $k^\mu$ is some (future-pointing) null vector field.
Taking into account Eq.~(\ref{phiprime}),
the NEC reduces to
\begin{equation}
\Xi = -\frac{Q^2}{p h^3} k^l k^l \ge 0 \ .
\end{equation}

For the choice of null vector field
\begin{equation}
k^\mu = \left(e^{-a}, \frac{e^{q/2}}{\sqrt{p}}, 0, \omega e^{-a},\omega e^{-a} \right) 
\end{equation}
we find, however,
\begin{equation}
\Xi = - Q^2\frac{e^{q}}{p^2 h^3} \le 0 \ .
\end{equation}
The dependence of $\Xi$ on the rotation parameter $c_\omega$ is discussed below.

\section{Wormhole solutions}
\subsection{Static wormhole solutions}

The static wormhole solutions can be given in analytical form.
For static wormholes $c_\omega=0$ and $q=a$. 
Thus the metric simplifies to
\begin{eqnarray}
ds^2 & = &  -e^{2a} dt^2 + p e^{-a}\left[dl^2 +(l^2+r_0^2) d\Omega_3^2\right] \ ,
\label{lineelstat}
\end{eqnarray}
where $d\Omega_3^2$ denotes the metric of a three-dimensional sphere.

In this case Eqs.~(\ref{eqa})
and (\ref{eqq}) are identical, 
\begin{equation}
\left(p (l^2+r_0^2)^{\frac{3}{2}} a'\right)' = 0 \ .
\label{stateq}
\end{equation}
With $p(l)$ given by Eq.~(\ref{solp}) the general solution is
\begin{equation}
a(l) = \frac{\mu}{r_0^2} 
      \arctan\left\{\frac{1-\sin z}{1+\sin z}\right\} +c_0 \  ,
\label{sola}
\end{equation}
where  $z=\arctan(l/r_0)$, and  $c_0$ and $\mu$ are integration constants.
Since we require asymptotically flat solutions in the limit $l\to \infty$,
the integration constant $c_0$ must vanish, $c_0=0$.
The constant $\mu$ determines the mass,  
\begin{equation}
\mu = \frac{8 G}{3\pi} M \ .
\end{equation}
For symmetric static wormholes the function $a$ is trivial, $a=0$.
Thus $\mu=0$, and the mass $M$ vanishes.

\begin{figure}[h!]
\begin{center}
\mbox{\hspace{0.2cm}
\subfigure[][]{\hspace{-1.0cm}
\includegraphics[height=.28\textheight, angle =0]{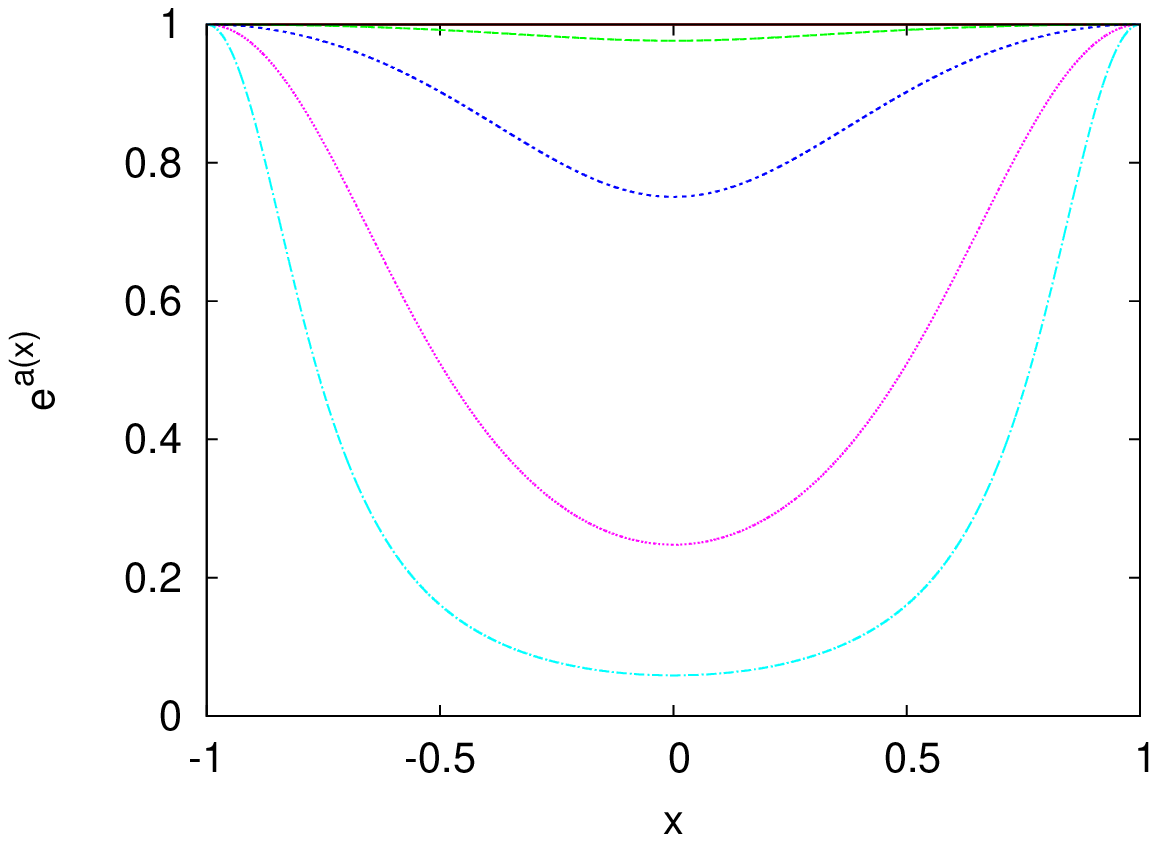}
\label{fig0a}
}
\subfigure[][]{\hspace{-0.5cm}
\includegraphics[height=.28\textheight, angle =0]{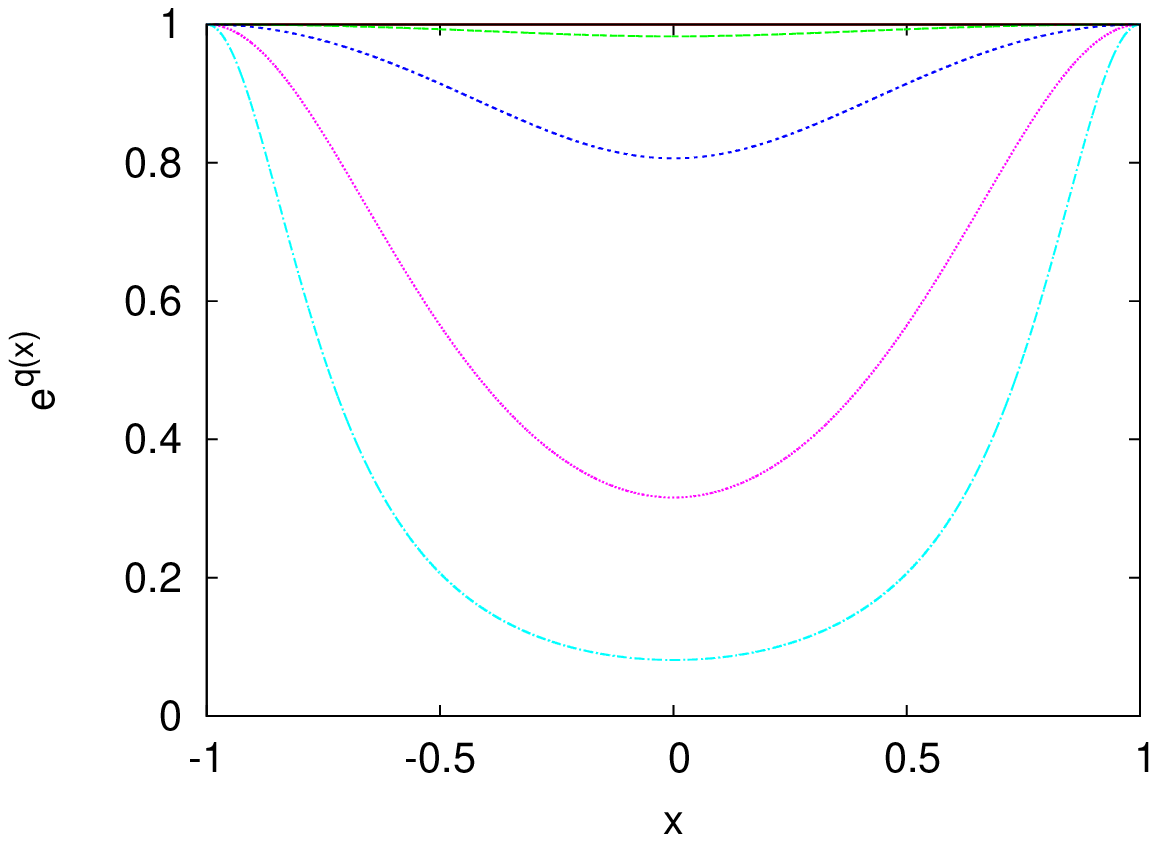}
\label{fig0b}
}
}
\mbox{\hspace{0.2cm}
\subfigure[][]{\hspace{-1.0cm}
\includegraphics[height=.28\textheight, angle =0]{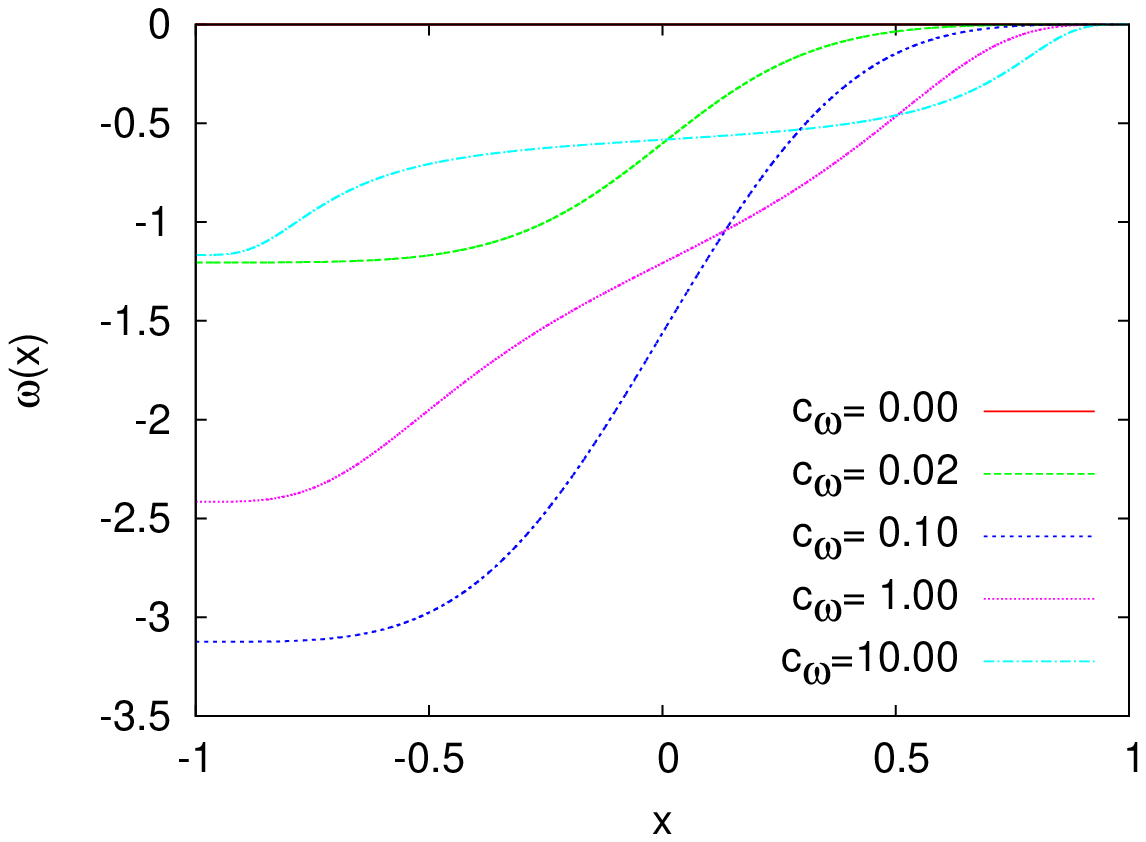}
\label{fig0c}
}
\subfigure[][]{\hspace{-0.5cm}
\includegraphics[height=.28\textheight, angle =0]{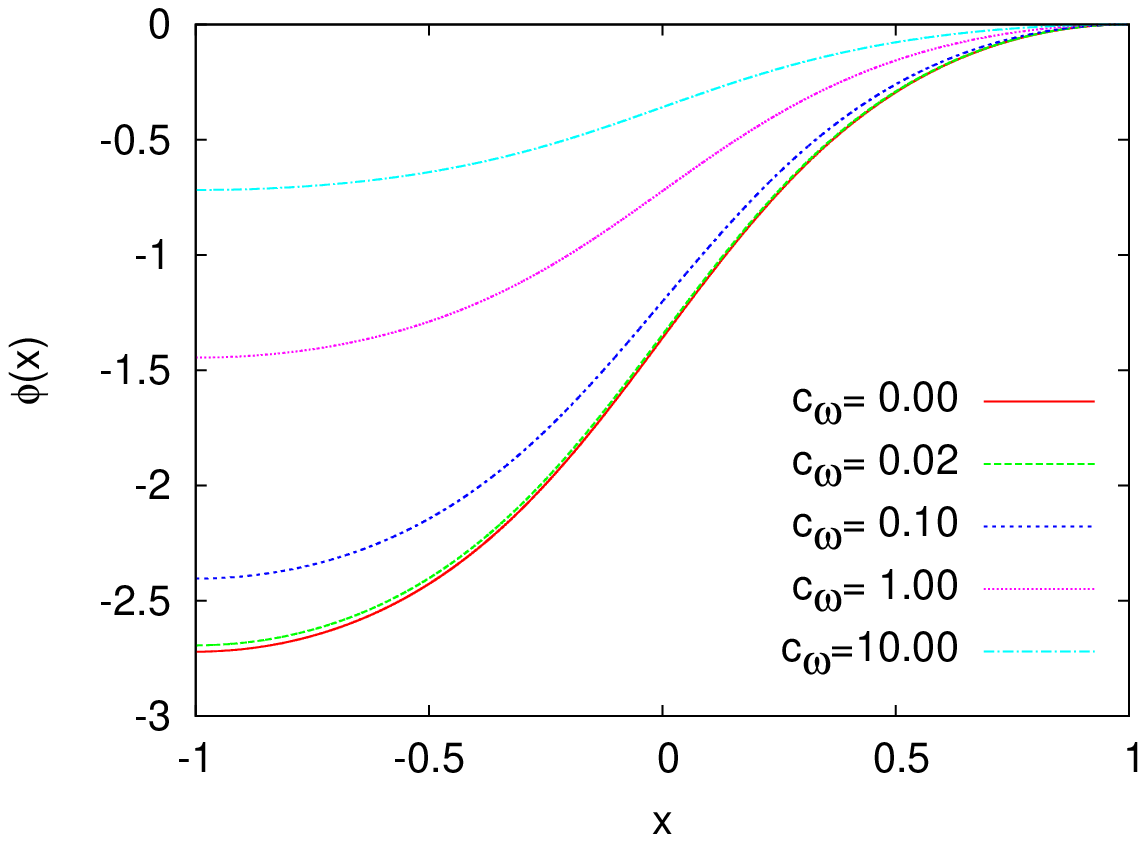}
\label{fig0d}
}
}
\end{center}
\vspace{-0.5cm}
\caption{The metric functions $e^a$ (a), $e^q$ (b) and $\omega$ (c) and 
the scalar function $\phi$ (d)
are shown for symmetric wormholes ($a_{-\infty}=0$) as functions of the  compactified coordinate $x$ 
for several values of the rotation constant $c_\omega$
and for throat radius $r_0=0.5$.
\label{Fig0}
}
\end{figure}

\begin{figure}[h!]
\begin{center}
\mbox{\hspace{0.2cm}
\subfigure[][]{\hspace{-1.0cm}
\includegraphics[height=.28\textheight, angle =0]{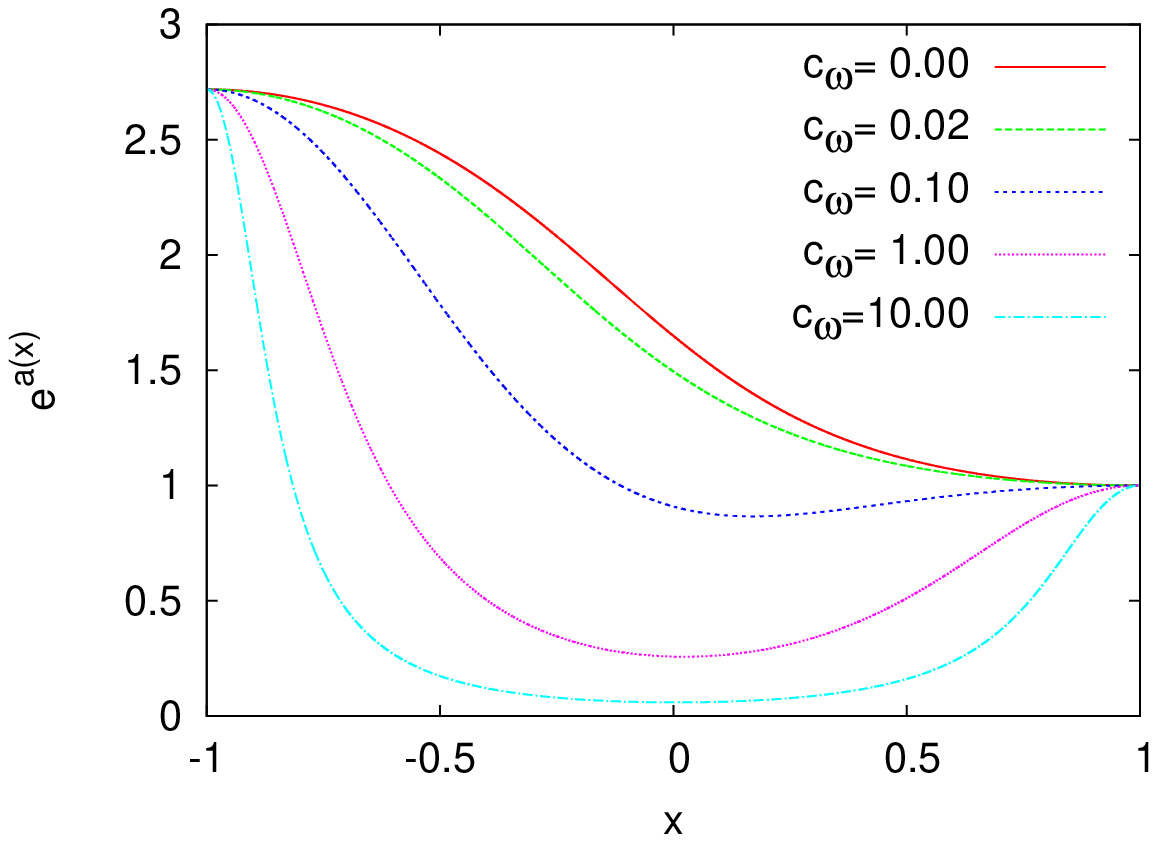}
\label{fig00a}
}
\subfigure[][]{\hspace{-0.5cm}
\includegraphics[height=.28\textheight, angle =0]{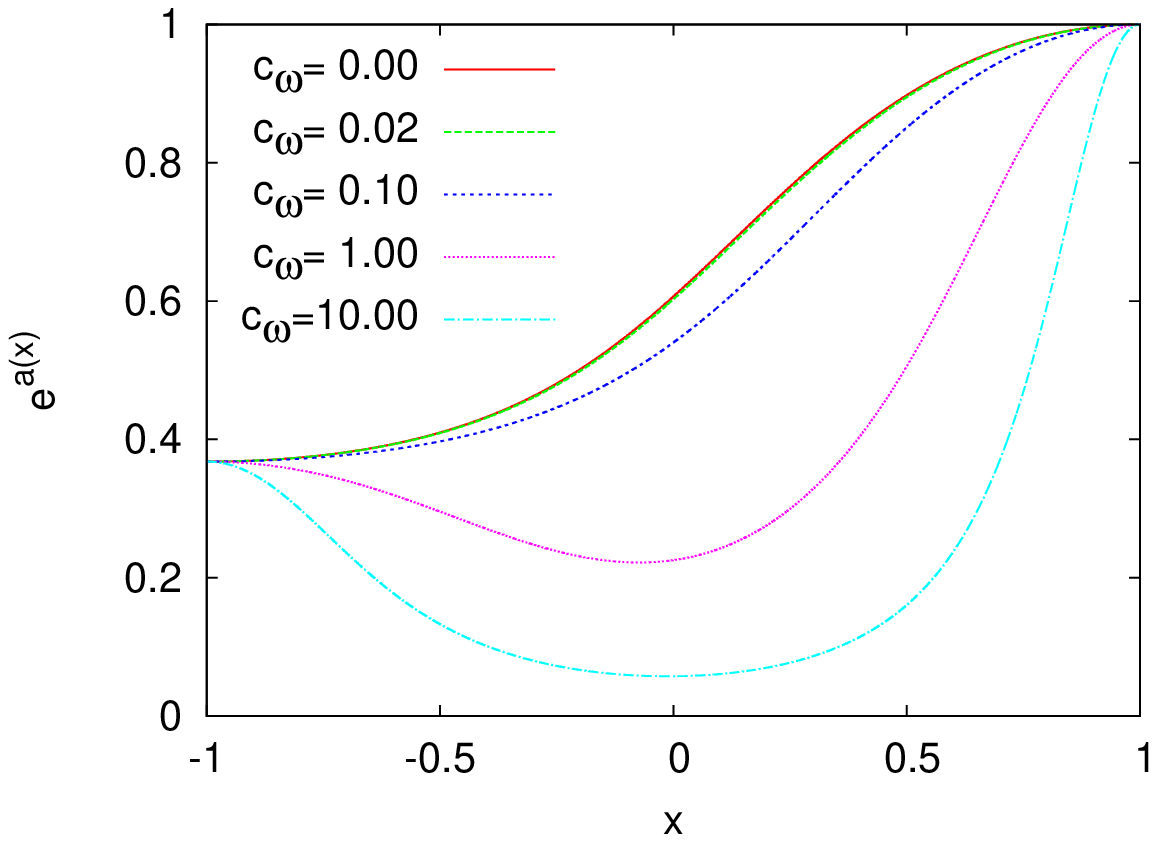}
\label{fig00b}
}
}
\mbox{\hspace{0.2cm}
\subfigure[][]{\hspace{-1.0cm}
\includegraphics[height=.28\textheight, angle =0]{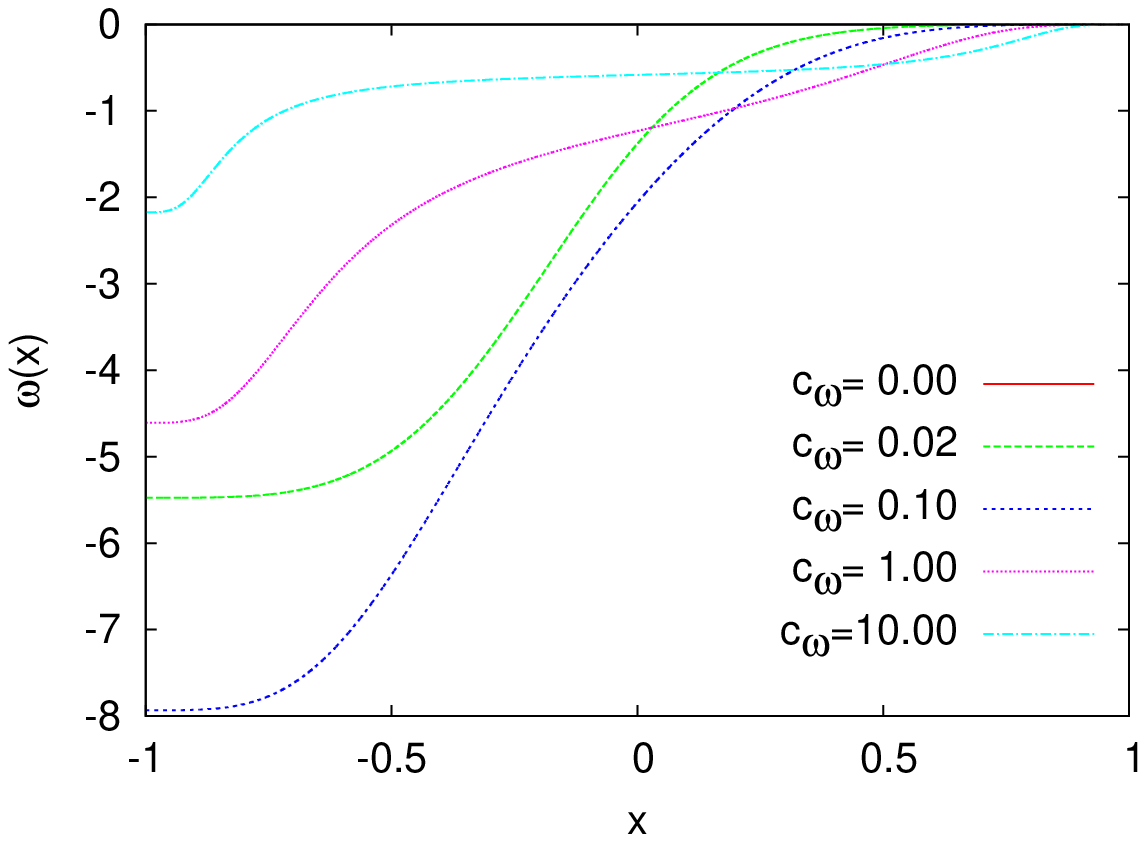}
\label{fig00c}
}
\subfigure[][]{\hspace{-0.5cm}
\includegraphics[height=.28\textheight, angle =0]{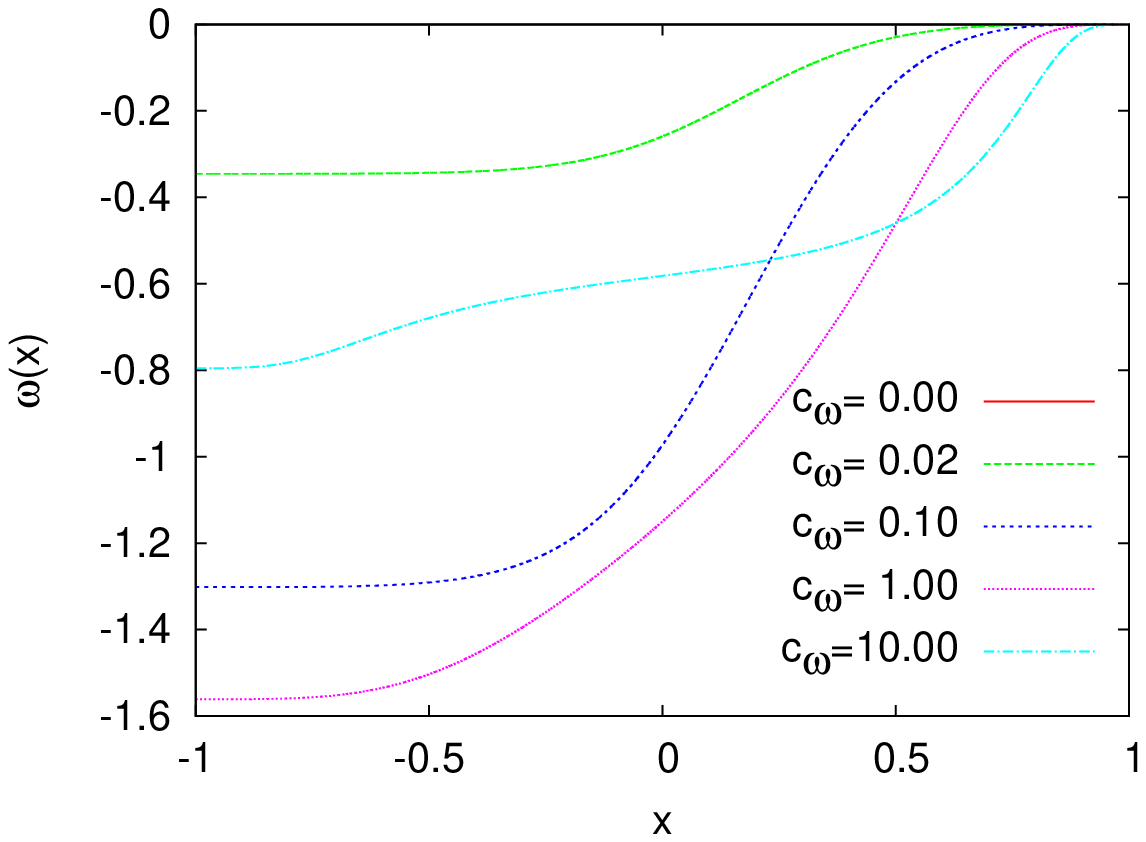}
\label{fig00d}
}
}
\mbox{\hspace{0.2cm}
\subfigure[][]{\hspace{-1.0cm}
\includegraphics[height=.28\textheight, angle =0]{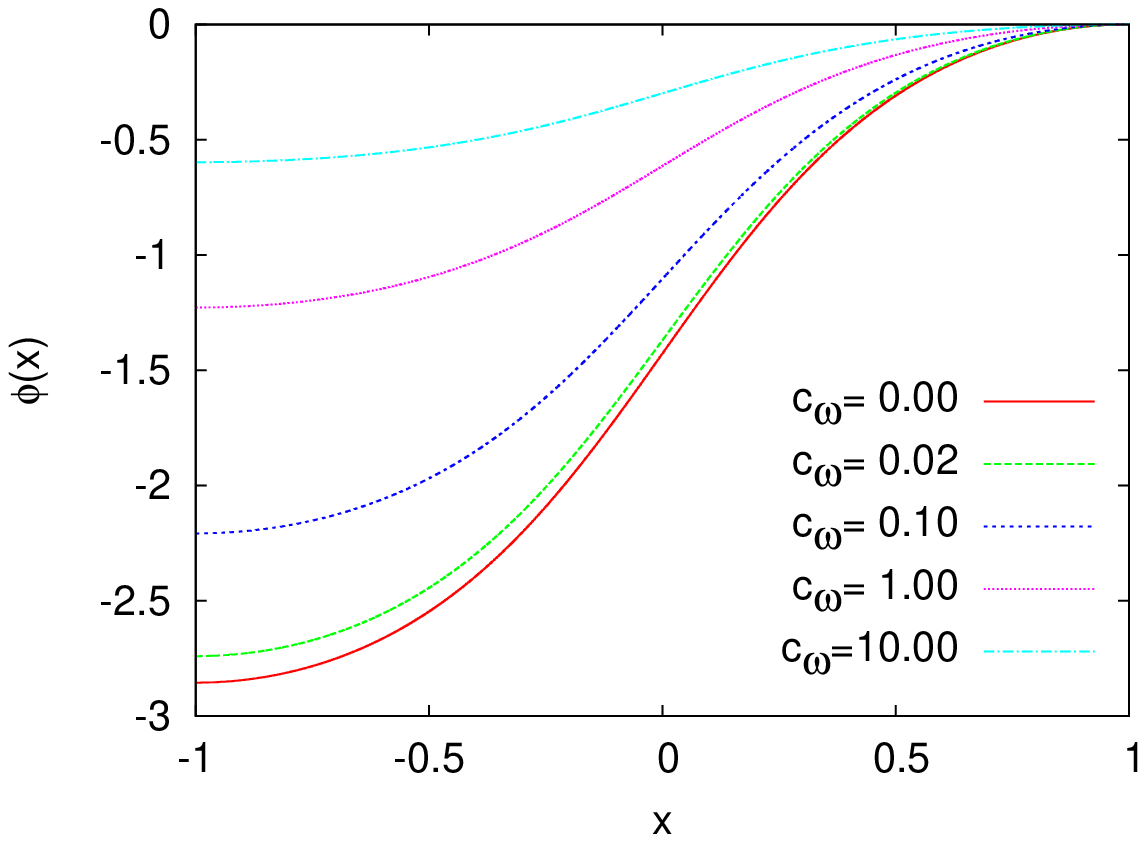}
\label{fig00e}
}
\subfigure[][]{\hspace{-0.5cm}
\includegraphics[height=.28\textheight, angle =0]{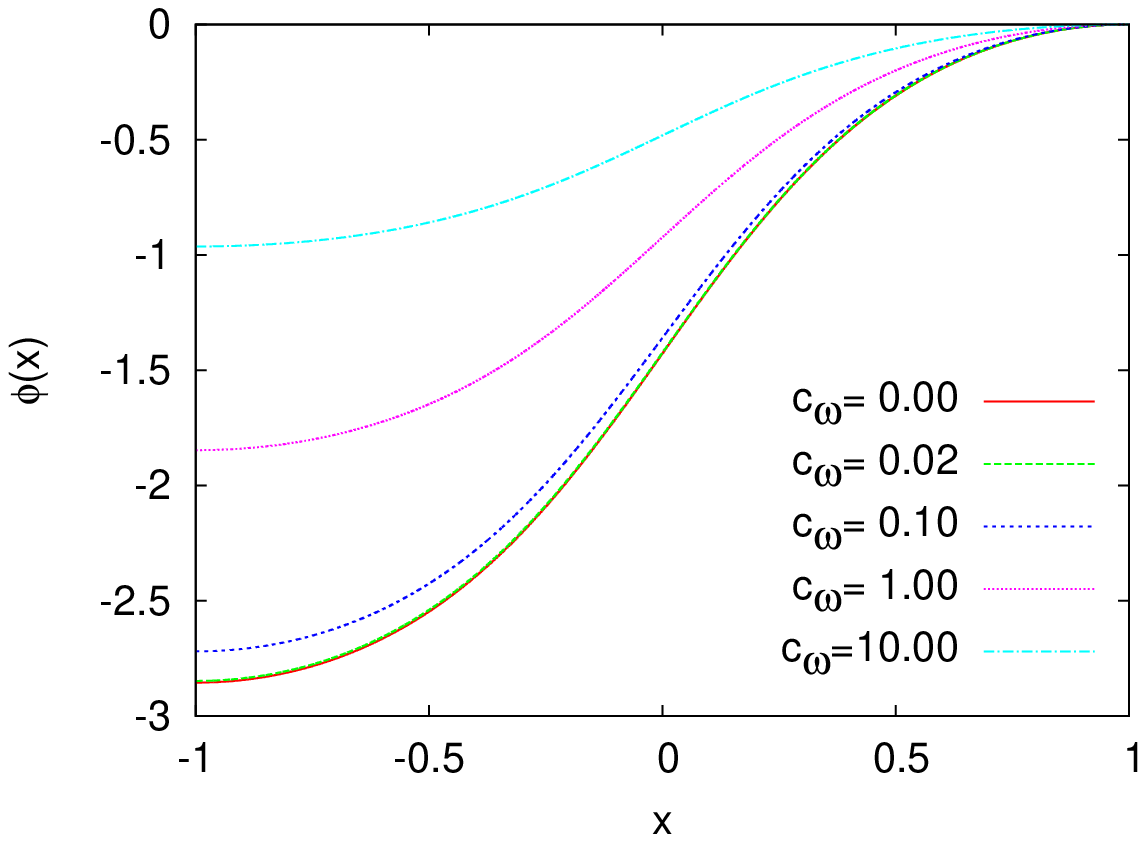}
\label{fig00f}
}
}
\end{center}
\vspace{-0.5cm}
\caption{The metric functions $e^a$ and $\omega$ and 
the scalar function $\phi$ 
are shown for non-symmetric wormholes (left column: $a_{-\infty}=1$,
right column: $a_{-\infty}=-1$) as functions of the  compactified coordinate $x$ 
for the same set of parameters as in Fig.~\ref{Fig0}.
\label{Fig00}
}
\end{figure}

For the scalar charge we find
\begin{equation}
4\pi G Q^2 = \frac{3}{4}\left(r_0^4+\mu^2\right) \ .
\end{equation}
The location of the throat is determined from
\begin{equation}
\sin z_0 = \frac{\sqrt{r_0^4+\mu^2}-r_0^2}{\mu} \ .
\end{equation}

The metric function $a$ and the scalar function $\phi$
are shown in Fig.~\ref{Fig0} for symmetric ($a_{-\infty}=0$)
wormhole solutions and in Fig.~\ref{Fig00} 
for non-symmetric ($a_{-\infty}=\pm 1$) wormhole solutions.

Since the static wormhole solutions represent
the starting solutions of the branches of rotating solutions
in the limit $J \to 0$,
their mass and their scalar charge
can be read off from Figs.~\ref{Fig1} in this limit.
Their stability will be discussed in the next section.

A generalization of this exact solution for any $D\geq 4$ is given in Appendix A.

\subsection{Stationary rotating wormhole solutions}

\subsubsection{Numerical scheme}

Stationary rotating wormhole solutions emerge smoothly from
the static ones when the angular momentum is slowly increased from zero.
An exact solution can be found in this case by treating $c_w$ as 
a small expansion parameter, see Appendix B.
To obtain the complete solutions, however, we have to resort to numerical methods.
We introduce the compactified coordinate $x$ via
\begin{equation}
l = r_0 \tan\left(\frac{\pi}{2} x\right) \ , \ \ \ \ 
-1 < x < 1 \ .
\end{equation}
At the boundaries the functions $a$ and $q$ must satisfy
the conditions
$a(-1) =  q(-1)$ and $a(1)= q(1)$.
We then solve
the system of ODEs with the boundary conditions
\begin{equation}
a(-1) =  q(-1) = 0 \ , \ \ a(1)= q(1) = 0 \ 
\end{equation}
to obtain symmetric wormholes, while the boundary conditions
$a(-1) =  q(-1) \ne 0$ lead to non-symmetric solutions (the approximate
form of the symmetric solutions in the asymptotic regions is given in Appendix B,
together with their expressions near $l=0$).

In the numerical calculations
a Newton-Raphson scheme is employed.
To obtain the branches of rotating solutions, we start from the corresponding static solutions
and then increase successively the parameter $c_\omega$,
while  we keep the throat parameter $r_0$ fixed.

\subsubsection{Metric and scalar functions}

Let us now consider the solutions themselves and their dependence on
the angular momentum.
In Fig.~\ref{Fig0} we exhibit the metric functions $a$, $q$, and $\omega$
and the scalar function $\phi$
for a set of symmetric wormhole solutions with increasing rotation parameter
$c_\omega$ and fixed throat parameter $r_0$. 
In the static case $c_\omega=0$, and the metric functions
are trivial. Only the scalar function $\phi$ is nontrivial and
assumes its smallest values.

In the rotating case,
with increasing rotation parameter $c_\omega$,
the even functions $a$ and $q$ show a monotonic decrease
in the central region.
But both remain very similar to each other.
The odd metric function $\omega$, on the other hand, 
shows a non-monotonic behaviour,
when the rotation parameter $c_\omega$ is varied.
The scalar function $\phi$ increases with 
increasing rotation parameter $c_\omega$,
tending to zero in the limit of large rotation.

The effect of choosing non-symmetric boundary conditions
is illustrated in Fig.~\ref{Fig00},
where we display
the metric functions $a$ and $\omega$
and the scalar function $\phi$
for $a_{-\infty}=\pm 1$ 
and the same set of rotation parameters $c_\omega$
and the throat parameter $r_0$.
Note that, the metric functions $a$ and $q$ behave very similarly
also in the non-symmetric case.

\subsubsection{Global charges}

We now turn to the discussion of the global charges of the wormholes.
For fixed throat parameter $r_0$ the mass $M$ increases
with increasing rotation parameter $c_\omega$,
while at the same time the scalar charge $Q$ decreases.
The increase of the mass due to rotation is expected and was observed before
in second order perturbation theory \cite{Kashargin:2008pk}.

To display the properties of the families of rotating wormhole solutions
we make use of the scaling symmetry of the equations.
For black holes one usually displays the scaled horizon area
versus the scaled angular momentum, 
where scaling is done with respect to the mass.
Since the wormholes may have a vanishing mass for a finite throat area, however,
we here choose to scale the mass $M$ and the angular momentum $J$
with respect to the throat area $A$.

\begin{figure}[h!]
\begin{center}
\mbox{\hspace{-0.2cm}
\subfigure[][]{\hspace{-1.0cm}
\includegraphics[height=.28\textheight, angle =0]{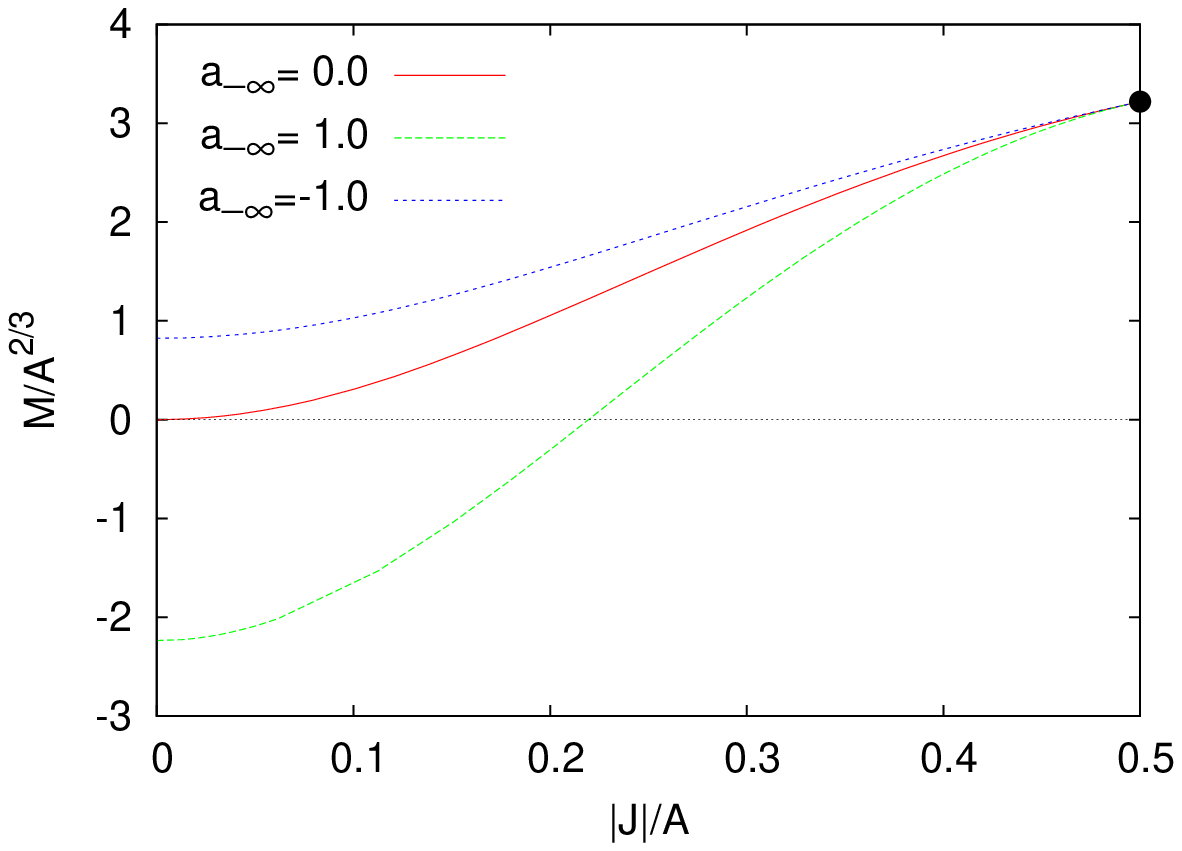}
\label{fig1a}
}
\subfigure[][]{\hspace{-0.5cm}
\includegraphics[height=.28\textheight, angle =0]{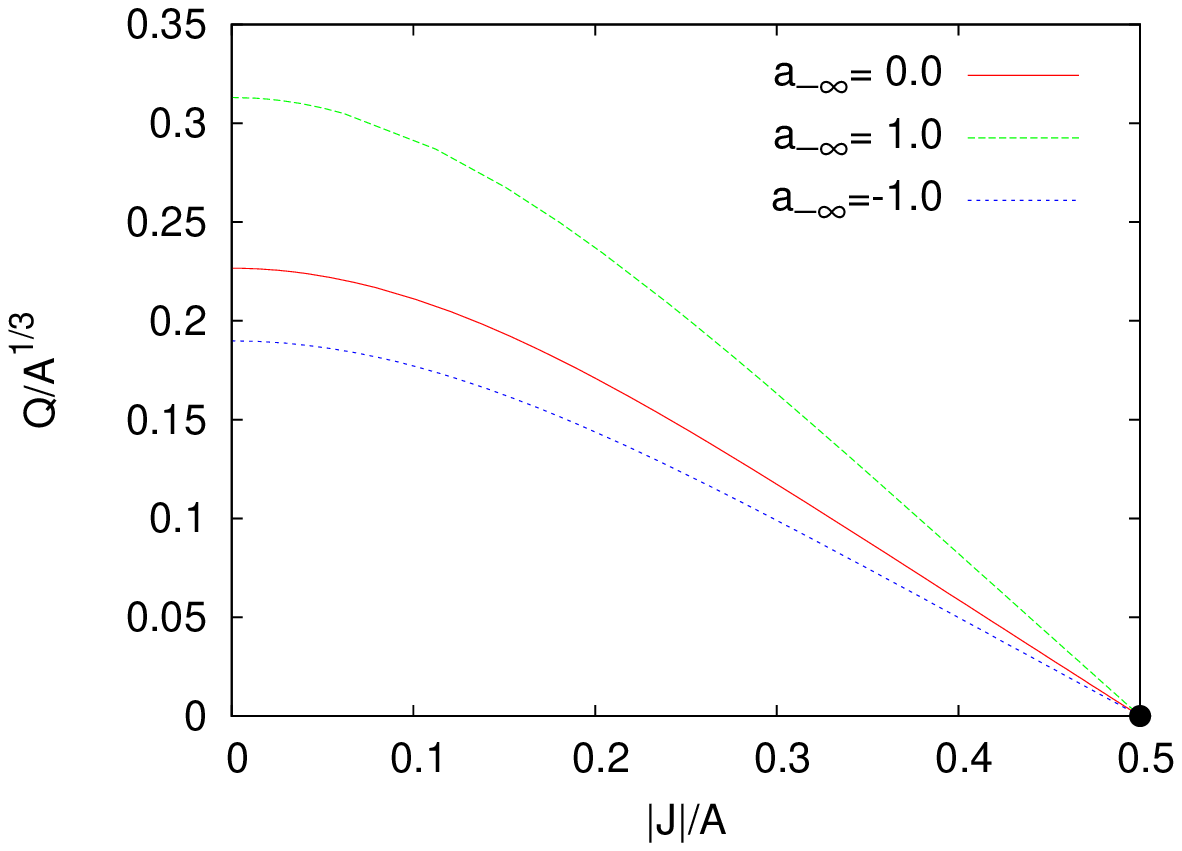}
\label{fig1b}
}
}
\end{center}
\vspace{-0.5cm}
\caption{The scaled mass $M/A^{2/3}$ (a)
and the scaled scalar charge $Q/A^{2/3}$ (b)
are shown as functions of the
scaled angular momentum $J/A$ for symmetric ($a_{-\infty}=0$) and 
non-symmetric ($a_{-\infty}=1, -1$) wormhole solutions.
The dots indicate the corresponding values 
of the extremal Myers-Perry black hole.
\label{Fig1}
}
\end{figure}

In Fig.~\ref{fig1a} we then show the scaled mass $M/A^{2/3}$ as a function of the
scaled angular momentum $J/A$.
Starting from the value of the scaled mass of the static solution 
for a given boundary value $a_{-\infty}$, the scaled mass increases monotonically
with increasing scaled angular momentum.
Interestingly, analogous to rotating black holes (in 4 and 5 dimensions),
the scaled angular momentum is bounded from above.
Moreover, as illustrated in the figure, this limiting extremal value
is the same for symmetric and non-symmetric wormholes.

In general, the scaled mass  $M/A^{2/3}$ shows a strong dependence on the
(symmetry breaking) boundary value $a_{-\infty}$,
increasing monotonically with increasing $a_{-\infty}$.
While this dependence is strongest in the static limit,
it disappears for the limiting extremal value
of the scaled angular momentum $J/A$.
This indicates that the same limiting solution is approached
in all these cases.
Note that, whereas  
the mass, the angular momentum and the area of the throat all diverge
for $c_\omega \to \infty$,
the scaled mass and scaled angular momentum assume finite values
in this limit.

The scaled scalar charge $Q/A$ exhibits a somewhat analogous dependence
on the scaled angular momentum $J/A$, as seen in Fig.~\ref{fig1b},
except that it decreases monotonically with increasing $J/A$.
Again, independent of the boundary value $a_{-\infty}$
the same value $Q/A=0$ is approached for the limiting extremal value
of the scaled angular momentum $J/A$.
Since in this limit the scalar charge vanishes one may expect
that the scalar field itself becomes trivial.
The limiting solution should therefore be a vacuum solution
in five dimensions with finite mass and equal angular momenta.

\begin{figure}[h!]
\begin{center}
\mbox{\hspace{0.2cm}
\subfigure[][]{\hspace{-1.0cm}
\includegraphics[height=.28\textheight, angle =0]{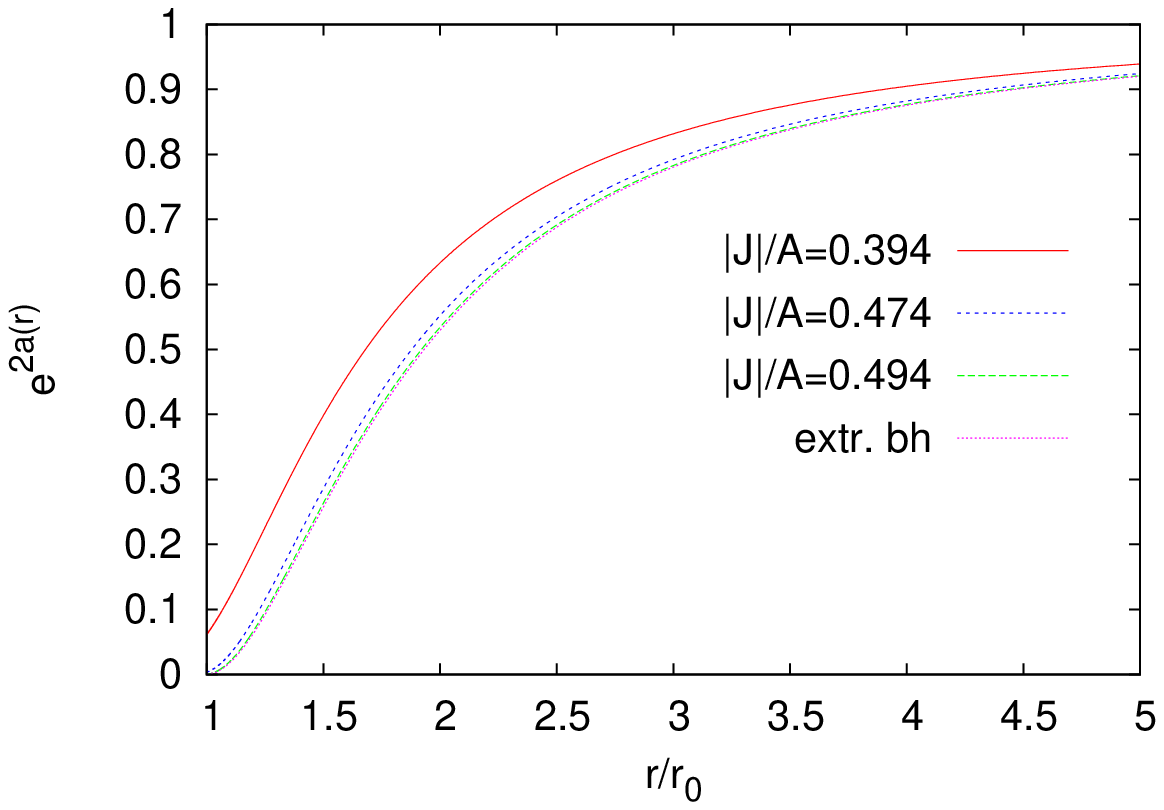}
\label{Fig2a}
}
\subfigure[][]{\hspace{-0.5cm}
\includegraphics[height=.28\textheight, angle =0]{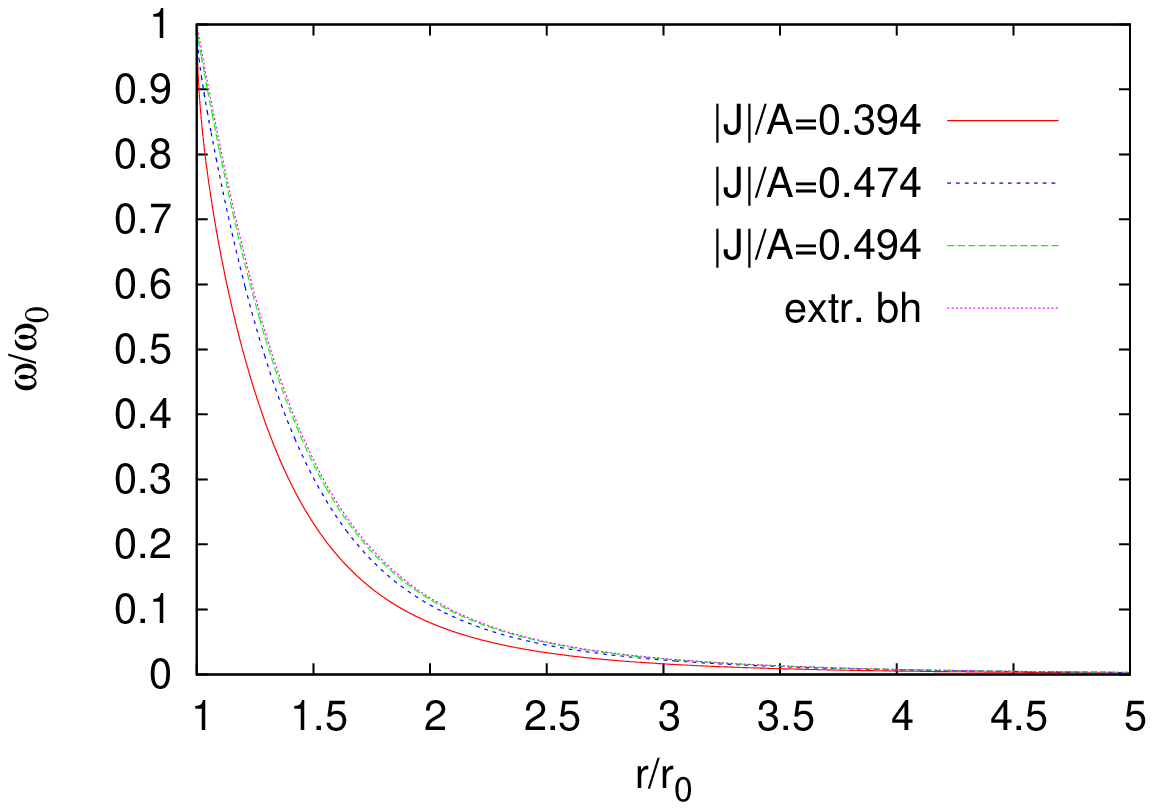}
\label{Fig2b}
}
}
\end{center}
\vspace{-0.5cm}
\caption{
The metric functions $e^{2a}$ (a) and $\omega$ (b) are shown
as functions of $r/r_0$ for several values 
of the scaled angular momentum $J/A$ together with the corresponding functions
of the extremal Myers-Perry black hole.
\label{Fig2}
}
\end{figure}

The family of rotating vacuum solutions that comes to mind is the  set of 
five-dimensional Myers-Perry black hole solutions with equal angular momenta,
and since we are looking for a limiting extremal solution we are bound to
consider the extremal black hole solution.
Indeed, evaluating the scaled mass and the scaled angular momentum
for the extremal Myers-Perry
black hole solution, by scaling with the horizon area $A_H$,
these values precisely match those limiting values of the rotating wormholes.
Clearly, the Smarr relation also matches in the limit.

Inspection of the wormhole solutions themselves then shows
that the metric functions
approach those of the extremal black hole solutions in the limit.
We demonstrate this for the metric functions $a$ and $\omega$
in Fig.~\ref{Fig2}, by exhibiting the wormhole functions 
for increasing values of the scaled angular momentum
close to the extremal limit,
and comparing them with the corresponding 
extremal black hole metric functions.
As noted above, the scalar field tends to zero in this limit.

\subsubsection{Geometry of the throat and ergoregion}

\begin{figure}[h!]
\begin{center}
\mbox{\hspace{0.2cm}
\subfigure[][]{\hspace{-1.0cm}
\includegraphics[height=.28\textheight, angle =0]{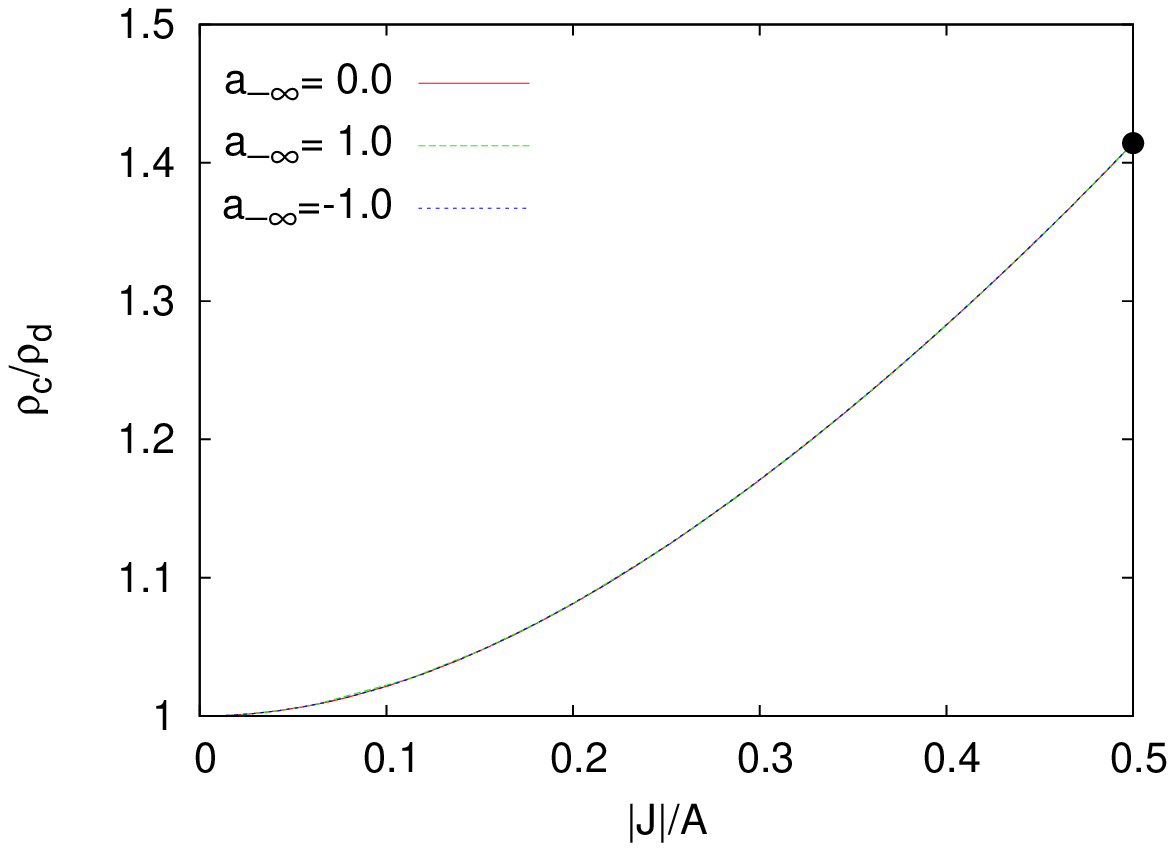}
\label{fig3a}
}
\subfigure[][]{\hspace{-0.5cm}
\includegraphics[height=.28\textheight, angle =0]{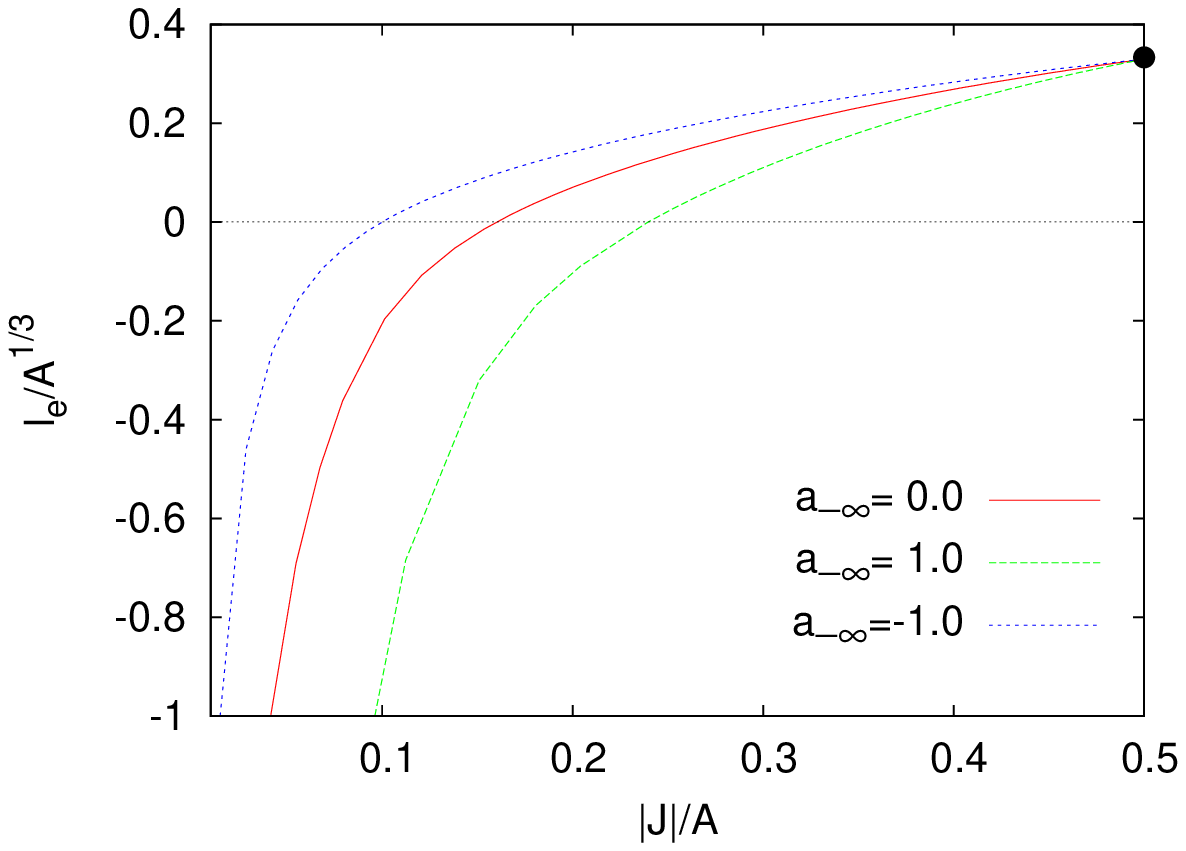}
\label{fig3b}
}
}
\end{center}
\vspace{-0.5cm}
\caption{The ratio $\rho_c/\rho_d$  (a) 
and the scaled coordinate $l_e/A^{1/3}$ of the location of the ergo hypersurface  (b) 
are shown as functions of the
scaled angular momentum $J/A$ for $a_{-\infty}=0$, $1$, and $-1$.
The dots indicate the corresponding values of the extremal Myers-Perry black hole.
\label{Fig3}
}
\end{figure}

Let us now consider the effect of the rotation on the
geometry of the throat.
While the static wormhole has a spherical throat, the throat
deforms when the wormhole rotates.
In order to demonstrate this deformation of the throat 
we show  in Fig.~\ref{fig3a} 
the ratio  of circumferences $\rho_c/\rho_d$ as a function of the 
scaled angular momentum.
In the limit of maximal rotation
this ratio assumes the value of the corresponding ratio
for the horizon of the extremal
Myers-Perry black hole.
Interestingly, the ratio seems to be independent of the 
boundary value $a_{-\infty}$.

As pointed out by Teo \cite{Teo:1998dp},
a rotating wormhole may possess an ergoregion.
In Fig.~\ref{fig3b} we show the scaled location $l_e/A^{1/3}$
of the ergo hypersurface
as a function of the scaled angular momentum $J/A$ for 
the boundary values $a_{-\infty}=0 $, $1$, and $-1$.
In the limit of maximal rotation
this ergo hypersurface agrees with the one of the extremal
Myers-Perry black hole.
For each $a_{-\infty}$ there is a value of the scaled angular momentum,
where the ergo hypersurface corresponds to the throat,
i.e., $l_e/A^{1/3} = 0$.
For smaller $J/A$, the ergo hypersurface is located 
at negative values of $l$, and thus resides in
the asymptotically rotating region.
For arbitrarily small but finite values of $J$, the location 
$l_e/A^{1/3}$ of the ergo hypersurface then tends to $-\infty$.
Thus the ergoregion disappears in the limit.
This agrees with the static solution which has no ergoregion.

\subsubsection{Null energy condition}

\begin{figure}[h!]
\begin{center}
\mbox{\hspace{0.2cm}
\subfigure[][]{\hspace{-1.0cm}
\includegraphics[height=.28\textheight, angle =0]{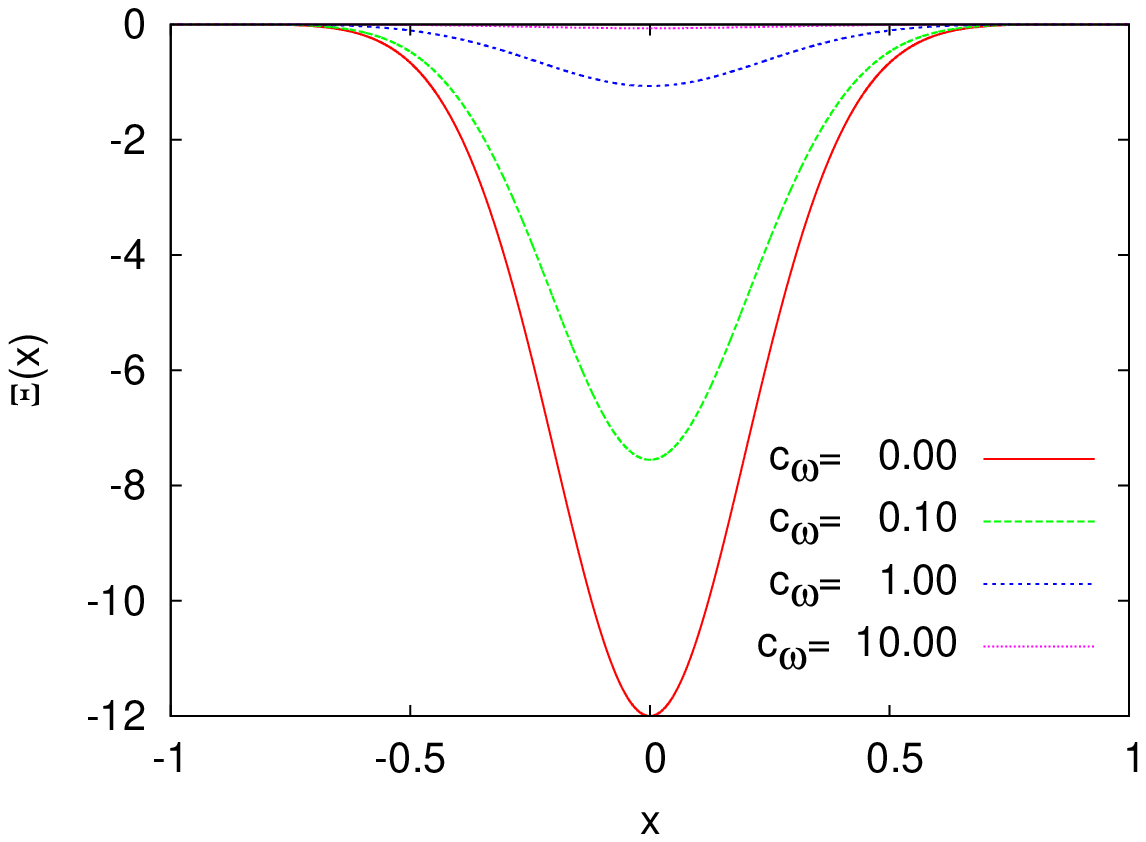}
\label{fig000a}
}
\subfigure[][]{\hspace{-0.5cm}
\includegraphics[height=.28\textheight, angle =0]{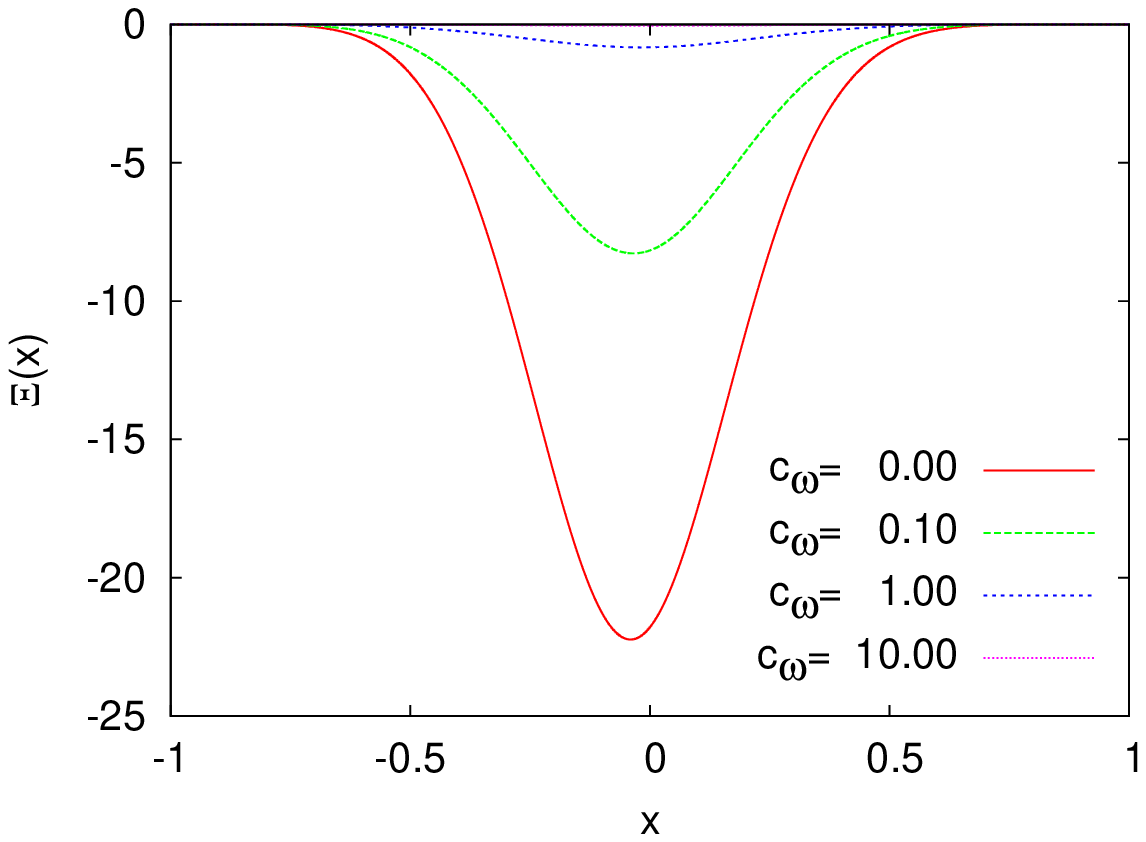}
\label{fig000b}
}
}
\end{center}
\vspace{-0.5cm}
\caption{The quantity $\Xi = - Q^2\frac{e^{q}}{p^2 h^3}$
is shown for symmetric  ($a_{-\infty}=0$) (a)
and non-symmetric ($a_{-\infty}=1$) (b) wormholes
as a function of the  compactified coordinate $x$ 
for several values of the rotation constant $c_\omega$
and for throat radius $r_0=0.5$.
The NEC would require $\Xi \ge 0$.
\label{Fig000}
}
\end{figure}

We have seen already that the NEC is violated for these five dimensional
wormholes. The quantity
$\Xi = - Q^2\frac{e^{q}}{p^2 h^3}$ 
is negative in the central parts of the spacetime.
To see the effect of the rotation on the violation of the NEC
we exhibit the quantity $\Xi(x)$ in Fig.~\ref{Fig000}
for the symmetric wormhole solutions of Fig.~\ref{Fig0} and 
the non-symmetric $a_{-\infty}=1$ solutions of Fig.~\ref{Fig00}.

We observe that for fixed $r_0$, the NEC violation
increases with increasing boundary value $a_{-\infty}$.
The NEC violation is maximal for the static solutions
and decreases with increasing rotation parameter $c_\omega$,
an effect seen already in second order perturbation theory \cite{Kashargin:2008pk}.
In the limit $c_\omega \to \infty$, the NEC violation disappears,
as it should for extremal Myers-Perry black holes.

\section{Stability}

\subsection{Perturbation equations}

Stability is vital for the traversability of wormholes.
Therefore we now turn to the 
study of the stability 
of the above wormhole solutions.
We start with the Ansatz for the line element 
and phantom field
\begin{eqnarray}
ds^2 & = &  -e^{2F_0} dt^2 + p e^{-F_0}\left\{
e^{F_0-F_2}\left[e^{F_1}dl^2 +h d\theta^2\right]
                   + e^{F_3-F_0}
		   h\left[ \sin^2\theta (d\varphi -F_\omega dt)^2
		         +\cos^2\theta (d\psi -F_\omega dt)^2\right]\right.
\nonumber\\
& &			\left.
			+\left(e^{F_0-F_2}- e^{F_3-F_0}\right)
			h \sin^2\theta\cos^2\theta(d\psi -d\varphi)^2
			\right\}
\ , \label{pertlineel}
\\
\Phi & = & \Phi(l,t) 
\ , \label{pertphi}
\end{eqnarray}
where $F_0, F_1, F_2, F_3$, and $F_\omega$ are functions of $l$ and $t$.
We substitute the Ansatz
in the Einstein and scalar field equations and introduce small
perturbations with harmonic time dependence,
\begin{equation}
F_0= a + \lambda H_0 e^{i \Omega t} \ , \ \  
F_1=     \lambda H_1 e^{i \Omega t} \ , \ \  
F_2= q + \lambda H_2 e^{i \Omega t} \ , \ \  
F_3= q + \lambda H_3 e^{i \Omega t} \ , \ \ 
F_\omega= \omega + \lambda H_\omega e^{i \Omega t} \ , \ \  
\Phi= \phi  + \lambda H_\phi e^{i \Omega t} \ , 
\end{equation}
where $H_0, H_1, H_2, H_3, H_\omega$, and $H_\phi$ are functions of $l$ only.
Expanding the Einstein and scalar field equations in $\lambda$,
the perturbation equations are then given by the terms of first order 
in $\lambda$.

{From} the $E_r^\vphi$ equation we obtain
\begin{equation}
H_\omega' = \frac{c_\omega e^{4a-q}}{p^2 h^{5/2}}\left[8H_0 + H_1 + H_2 - 3 H_3\right] \ .
\label{hop}
\end{equation}
The perturbation equation for the scalar field suggests to impose the gauge
\begin{equation}
H_1 = H_3-H_2\ .
\label{gauge}
\end{equation}
This implies that
$H_\phi = 0$ for unstable modes. 
To show this we multiply 
the ODE for $H_\phi$ by $H_\phi$,
\begin{equation}
H_\phi \left( p h^{3/2} H_\phi'\right)' 
+\Omega^2 p^2 h^{3/2} e^{-2a-q} H_\phi^2 = 0 \ ,
\label{eqhphi}
\end{equation}
where the gauge choice Eq.~(\ref{gauge}) has been taken into account.
Intergration yields 
\begin{equation}
 \left[ p h^{3/2}H_\phi H_\phi'\right]_{-\infty}^{\infty} = 
 \int_{-\infty}^{-\infty} \left\{p h^{3/2} (H_\phi')^2 
                     -\Omega^2 p^2 h^{3/2} e^{-2a-q} H_\phi^2\right\}dl \ ,
\label{inteqhphi}
\end{equation}
where integration by parts was used. 
Since the lhs vanishes, the rhs has to vanish as well.
But for unstable modes $ -\Omega^2$ is positive. 
Hence the integrand is non-negative.
Consequently, the integral only vanishes if $H_\phi=0$ for all $l$.

This leaves us with three equations for $H_2, H_3$, and $H_0$.
We find it convenient to introduce
\begin{equation}
H_0 =  G_2 + G_0 \ , \ \ H_2 = -G_2 \ , \ \ H_3 = 2 G_0 +G_1 \ .
\nonumber
\end{equation}
The resulting perturbation equations form a system of homogeneous ODEs
\begin{eqnarray}
\frac{1}{\xi}(\xi G_0')' & = & V_{00} G_0 +V_{01} G_1+V_{02} G_2 -\Omega^2\left(G_0+G_1\right)e^{-2a-q} p \ ,
\label{eqG0}\\
\frac{1}{\xi}(\xi G_1')' & = & V_{10} G_0 +V_{11} G_1+V_{12} G_2 -\Omega^2G_1 e^{-2a-q} p \ ,
\label{eqG1}\\
\frac{1}{\xi}(\xi G_2')' & = & V_{20} G_0 +V_{21} G_1+V_{22} G_2 -\Omega^2G_2 e^{-2a-q} p \ ,
\label{eqG2}
\end{eqnarray}
where $\xi = p h^{3/2}$ 
and the $V_{ab}$ are functions of the unperturbed solutions,
\begin{eqnarray}
V_{00} & = & - \frac{8 ( 2 e^{2a} - e^{2q}) e^q h^3 p^3 - c_\omega^2 e^{6a} }{  h^4 p^3}e^{-2a-q} \label{eqv00} \ , \\
V_{01} & = & - \frac{16(   e^{2a} - e^{2q}) e^q h^3 p^3 + c_\omega^2 e^{6a} }{2 h^4 p^3}e^{-2a-q} \label{eqv01} \ , \\
V_{02} & = & -2\frac{   4( e^{2a} + e^{2q}) e^q h^3 p^3 - c_\omega^2 e^{6a} }{  h^4 p^3}e^{-2a-q} \label{eqv02} \ , \\
V_{10} & = &  2\frac{   4(4e^{2a} + e^{2q}) e^q h^3 p^3 - c_\omega^2 e^{6a} }{  h^4 p^3}e^{-2a-q} \label{eqv10} \ , \\
V_{11} & = &   \frac{8 (2  e^{2a} - e^{2q}) e^q h^3 p^3 + c_\omega^2 e^{6a} }{  h^4 p^3}e^{-2a-q} \label{eqv11} \ , \\
V_{12} & = &  4\frac{   2(2e^{2a} + e^{2q}) e^q h^3 p^3 - c_\omega^2 e^{6a} }{  h^4 p^3}e^{-2a-q} \label{eqv12} \ , \\
V_{20} & = &  8\frac{     2e^{2a} - e^{2q} }{h} e^{-2a}  \label{eqv20} \ , \\
V_{21} & = &  8\frac{      e^{2a} - e^{2q} }{h} e^{-2a}  \label{eqv21} \ , \\
V_{22} & = &  8\frac{      e^{2a} + e^{2q} }{h} e^{-2a}  \label{eqv22} \ .
\end{eqnarray}

\subsection{Unstable modes}

We first note 
that in the static case the ODEs possess a zero mode, 
\begin{equation}
G_0 =- 2G_2 \ , \ \ \ 
G_1 = 3 G_2 \ , \ \ \
G_2 = const. \ ,
\end{equation}
which is, however, not normalizable. 
In this case the perturbed line element is
\begin{equation}
ds^2  =   -dt^2 + p e^{\lambda}\left\{
dl^2 +h\left[ d\theta^2 + \sin^2\theta d\phi^2+\cos^2\theta d\psi^2\right]
\right\} \ ,
\label{pertlineelzero}
\end{equation}
where $\lambda$ is considered small.

To obtain normalizable modes,
we solve the system of ODEs with
a set of appropriate boundary conditions. In particular, we choose
\begin{equation}
G_0(-\infty) = G_1(-\infty)= G_2(-\infty)= 0 \ , \ \ 
G_0( \infty) = G_1( \infty)= G_2( \infty)= 0 \ , \ \ G_2(0) = 1 \ ,
\label{bcpert}
\end{equation}
where the last condition ensures that the solutions are nontrivial.
However, since the total order of the system of ODEs (i.e., $3\times 2 = 6$) 
has to match the number of boundary conditions,
we introduce an auxilliary first order ODE ${\Omega^2}'=0$. 
Then the correct value of $\Omega^2$ 
adjusts itself in the numerical procedure.

\begin{figure}[h!]
\begin{center}
\mbox{\hspace{0.2cm}
\subfigure[][]{\hspace{-1.0cm}
\includegraphics[height=.28\textheight, angle =0]{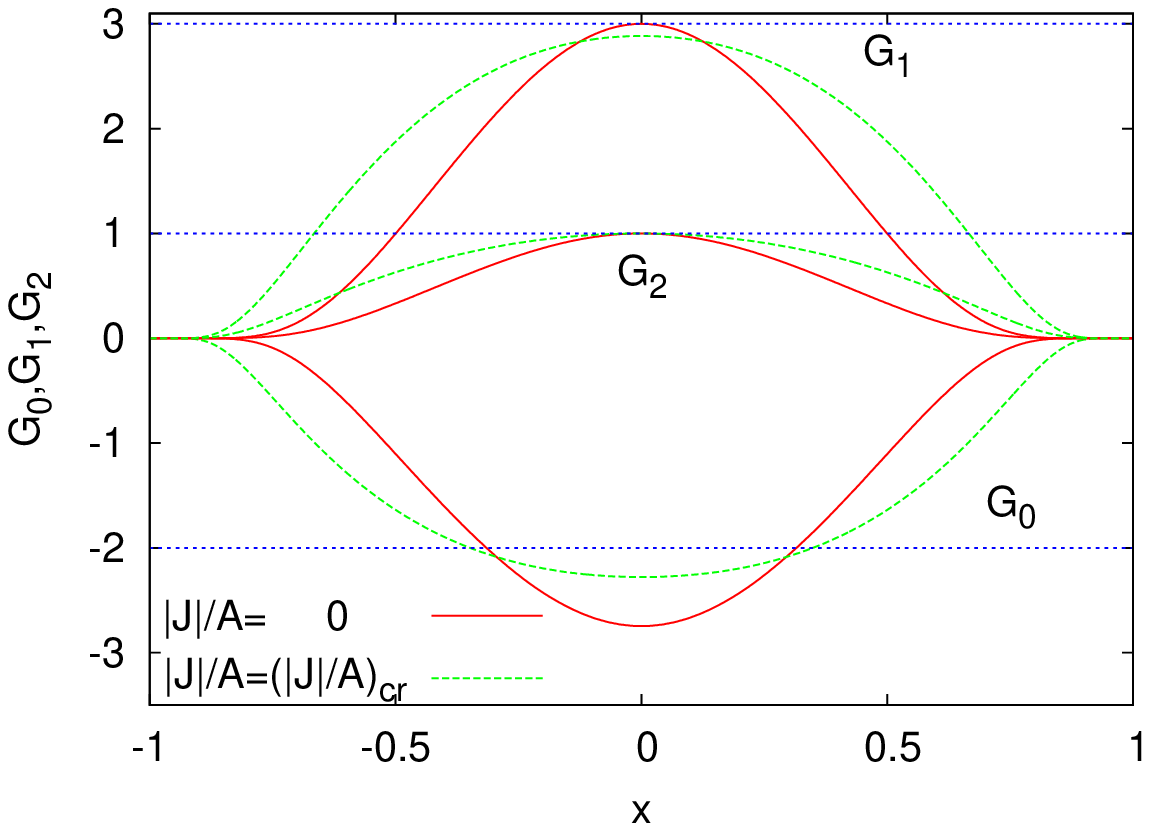}
\label{fig4a}
}
\subfigure[][]{\hspace{-0.5cm}
\includegraphics[height=.28\textheight, angle =0]{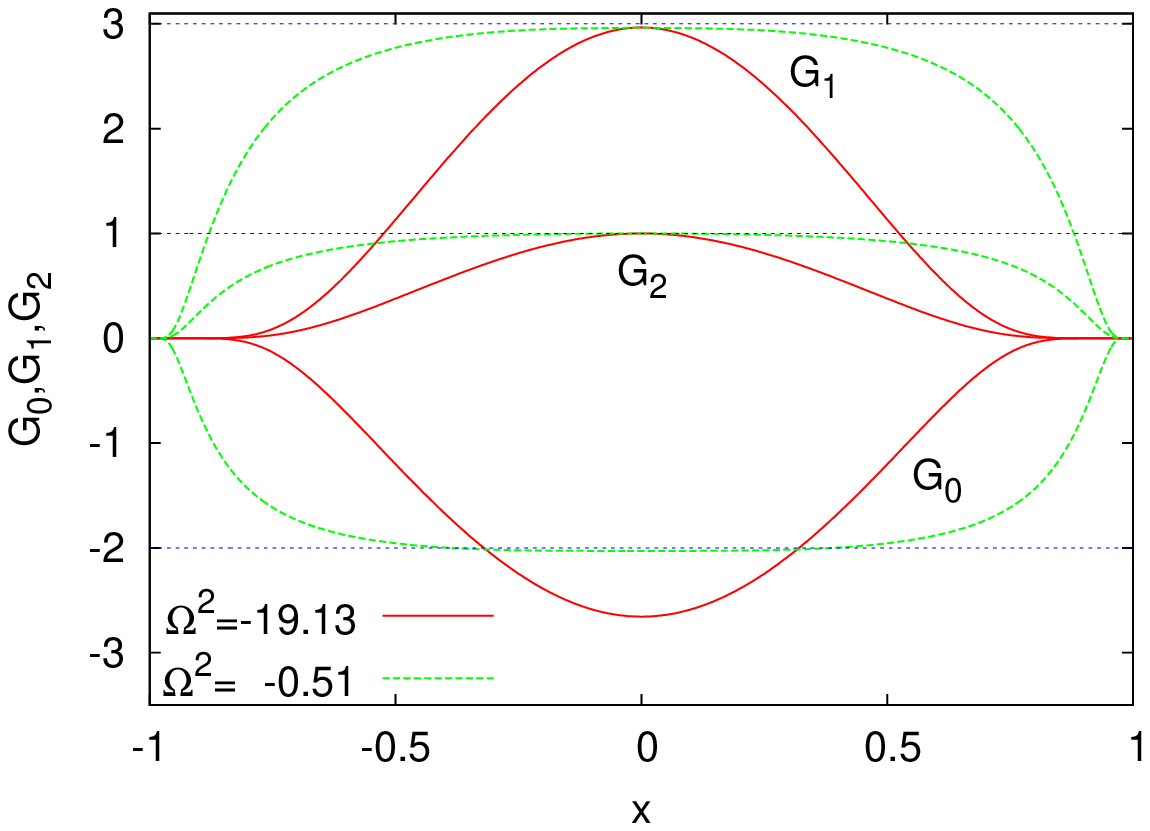}
\label{fig4b}
}
}
\end{center}
\vspace{-0.5cm}
\caption{The perturbation functions $G_0$, $G_1$, and $G_2$
are shown as functions of the   compactified coordinate $x$ for 
symmetric wormholes with throat parameter $r_0=0.5$:
(a) the single unstable mode of the static wormhole,
and the single unstable mode of the rotating wormhole at the critical
angular momentum of the bifurcation;
(b) the two unstable modes of the rotating wormhole at the
angular momentum $J/A=0.05475$.
The straight lines show the
non-normalizable zero mode of the static wormhole for comparison.
\label{Fig4}
}
\end{figure}

\begin{figure}[h!]
\begin{center}
\mbox{\hspace{0.2cm}
\includegraphics[height=.28\textheight, angle =0]{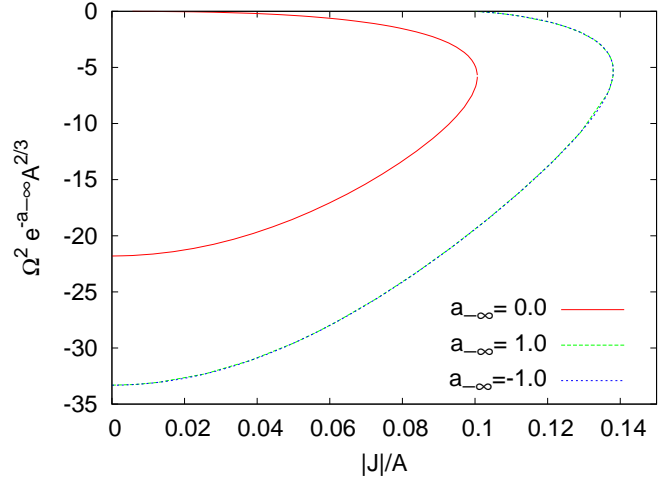}
}
\end{center}
\caption{The scaled eigenvalue
$\Omega^2 e^{-a_{-\infty}}A^{2/3}$ is shown as a function of the scaled angular momentum
$J/A$  for $a_{-\infty}=0,1,-1$.
\label{Fig5}
}
\end{figure}

Our numerical analysis of the equations shows that,
as expected, there is a normalizable unstable mode in the static case
(see also \cite{Shinkai:2013}).
It is exhibited in Fig.~\ref{fig4a}, where the functions
$G_0$, $G_1$, and $G_2$ are shown versus the compactified coordinate $x$
for a symmetric wormhole with throat parameter $r_0=0.5$.
The corresponding 
scaled eigenvalue is exhibited in Fig.~\ref{Fig5}.
For non-symmetric static wormholes, 
the scaled eigenvalue decreases with increasing $a_{-\infty}$, thus the instability increases.
The respective functions are then no longer symmetric.

The interesting issue is, however, how the rotation influences 
the stability of the wormholes.
Let us therefore now consider the rotating case.
Here we encounter a surprise.
The non-normalizable zero mode of the static wormholes
turns into a normalizable unstable mode of the
rotating wormholes.
With increasing angular momentum, this eigenvalue decreases 
and thus the mode becomes increasingly unstable.
On the other hand, as hoped for, the eigenvalue of the
unstable mode originating from the unstable mode of the static solutions increases
with increasing angular momentum, 
and thus this mode seems to tend towards stability.

For a typical intermediate value of the angular momentum
$J$,
the functions $G_0$, $G_1$, and $G_2$ 
for both unstable modes are shown in Fig.~\ref{fig4b}.
The functions for the lower value of $\Omega$ correspond to the
mode originating from the unstable mode of
the static solution.
The functions for the higher value of $\Omega$, in contrast,
represent the unstable mode associated with the
zero mode of the static solution.
This origin is still clearly recognizable, since the two sets of functions
still agree well in a large region around the throat.

Interestingly, however,
by increasing the angular momentum further, 
the two unstable modes get closer and finally merge
at a critical value $J_{\rm cr}$
of the angular momentum.
The perturbation functions of this critical solution,
where the bifurcation takes place,
are shown in Fig.~\ref{fig4a} for symmetric wormholes
with throat parameter $r_0=0.5$.

The dependence of the scaled eigenvalues of the two unstable modes
on the scaled angular momentum is exhibited in Fig.~\ref{Fig5},
both for symmetric wormholes ($a_{-\infty}=0$)
and non-symmetric wormholes ($a_{-\infty}=\pm 1$).
We note that when scaling the eigenvalue also with the metric
function $e^{-a_{-\infty}}$, the correspondingly
scaled curves depend only on the magnitude of $a_{-\infty}$.
We have therefore included this factor in the scaling.
The critical value $J_{\rm cr}$ of the angular momentum 
depends only on the magnitude of the
boundary value $a_{-\infty}$ and increases with increasing
$|a_{-\infty}|$.
Thus for non-symmetric wormholes the instability
persists to faster rotation.

For symmetric wormholes the critical value of the scaled
angular momentum is 
$(|J|/A)_{\rm cr} = 0.10063$,
whereas for non-symmetric wormholes with
boundary value $|a_{-\infty}|=1$ it is
$(|J|/A)_{\rm cr} = 0.1381$.
Beyond $J_{\rm cr}$ we do not find any further
unstable modes within the type of modes investigated.
We conclude that 
if no new unstable modes of a different type appear -
and we see no reason why new unstable modes should arise -
rotation may stabilize wormholes, if they rotate 
sufficiently fast.
If from a physical point of view the unstable mode
would tend to shrink the size of the throat, 
then beyond the critical value $(|J|/A)_{\rm cr}$
the centrifugal force would be strong enough
to inhibit such a shrinkage.

\section{Conclusions}

Employing a massless phantom field
we have constructed wormholes in five spacetime dimensions.
The static solutions have been given in closed form.
For the rotating wormholes we have considered only
cohomogeneity-1 solutions.
Here the two angular momenta - each associated with 
rotation in one of the two orthogonal planes - have equal magnitude,
$J_1=J_2=J$.
The rotating solutions have been obtained by numerical integration.

For a given size of the throat, we find a branch of rotating wormholes
starting at the corresponding static solution and ending in a
limiting extremal solution of maximal angular momentum $J$.
Interestingly, this limiting solution corresponds to
the extremal Myers-Perry black hole,
whose horizon has the same size as the wormhole throat.
The wormhole solutions satisfy a Smarr-like relation, which 
- for symmetric wormholes - is analogous
to the Smarr relation of extremal black holes,
where the angular velocity of the horizon is replaced by the
angular velocity of the throat.

With increasing angular momentum the throat becomes increasingly deformed.
For fixed horizon size this deformation depends only on the angular momentum
and is the same for symmetric and non-symmetric wormholes.
The violation of the NEC is strongest for static wormholes
and decreases with increasing angular momentum.
It decreases to zero when the limiting configuration is approached.
This is expected,
since the Myers-Perry solution has no phantom field
and no NEC violation.

The stability analysis of the wormhole solutions
has revealed surprising features.
A non-normalizable zero mode of the static solutions
turns into a normalizable unstable mode, when the  wormhole solutions
start to rotate.
Its frequency then decreases further with increasing angular momentum.
On the other hand,  the frequency of the unstable mode
that evolves from the unstable mode of the static solutions,
increases with increasing angular momentum.
Intriguingly,
at a critical value of the angular momentum $J_{\rm cr}$ the two unstable
modes then bifurcate and disappear.
We do not find unstable modes beyond $J_{\rm cr}$. 
This indicates that rotation may stabilize wormholes.
It would be interesting to apply the numerical methods of
\cite{Shinkai:2013} to study the time evolution of these
rotating wormholes.

The next step will be to construct self-consistent wormholes
in four dimensions, and to study their stability.
Representing a cohomogeneity-2 problem, 
this analysis will be much more involved.
The current results, however, give us hope as to be able
to find stable rotating wormholes.
Stability will also be essential in our envisaged study 
of rotating stars with wormholes at their cores,
a problem previously considered only for static stars
\cite{Dzhunushaliev:2011xx,Dzhunushaliev:2012ke,Dzhunushaliev:2013lna}.
Further astrophysically motivated studies may
include the study of the shadow of self-consistent rotating
wormholes \cite{Nedkova:2013msa}.

The static wormhole solutions can be straightforwardly generalized to
$D>5$ dimensions, where they can be obtained in analytical form
(see \cite{Shinkai:2013} and Appendix A).
The rotating solutions on the other hand will need numerical calculations.
Cohomogeneity-1 solutions are only present in odd dimensions.
They should be obtainable analogously to the ones presented here
\cite{Kunz:2006eh}.
Rotating wormholes in even dimensions, in contrast, 
represent at least a cohomogeneity-2 problem, and thus
sets of partial differential equations would need to be  solved.

Finally, based on our results presented here, we would like to speculate,
that in analogy to rotating black holes in $D>5$ dimensions
\cite{Emparan:2008eg},
which possess no extremal limit, when one of the angular momenta vanishes, 
the same may hold true for rotating wormholes.
In such a case, the throat might deform strongly,
giving rise to Gregory-Laflamme type instabilities.
Then a similar phase structure could arise
for higher dimensional rotating wormholes
as for black holes, allowing for wormholes with ringlike
throats and for composite wormholes.

\section*{Acknowledgements}

We gratefully acknowledge support by the German Research Foundation  
within the framework of the DFG Research Training Group 1620 
{\it Models of gravity}
as well as support by the Volkswagen Stiftung. 
VD and VF 
gratefully acknowledge a grant in fundamental research in natural sciences 
by the Ministry of Education and Science of Kazakhstan 
for the support of this research. 
They also would like to thank the Carl von Ossietzky University of Oldenburg 
for hospitality while this work was carried out.

\begin{appendix}
\section{Static wormholes in $D$ dimensions}
\setcounter{equation}{0}

Static wormholes in $D$ dimensions with a massless phantom field
have been considered in \cite{Shinkai:2013} by
employing, however, a different line element.

To obtain the static solutions in $D$ dimensions we choose the line element
\begin{equation}
ds^2 = -e^{(D-3)a} dt^2 +e^{-a}p\left[ dl^2 +(l^2+r_0^2) d\Omega_{D-2}^2\right] \ ,
\end{equation}
where $d\Omega_{D-2}^2$ denotes the metric of the $(D-2)$-dimensional sphere.
This Ansatz leads to the set of ODEs
\begin{eqnarray}
0 & = & p''+ \frac{2D-5}{l^2+r_0^2} l p' +2\frac{(4-D)r_0^2}{(l^2+r_0^2)^2} p +
       \frac{D-5}{2}\frac{p'^2}{p} \ ,
\label{Dpeq}\\  
0 & = & a''  + a'\left[(D-2) \frac{l}{l^2+r_0^2} + \frac{D-3}{2}\frac{p'}{p}\right] \ ,
\label{Daeq}\\  
0 & = & \phi'  + p^{\frac{3-D}{2}}(l^2+r_0^2)^{\frac{2-D}{2}}Q \ .
\label{Dphieq}
\end{eqnarray}

\begin{figure}[h!]
\begin{center}
\mbox{\hspace{0.2cm}
\includegraphics[height=.28\textheight, angle =0]{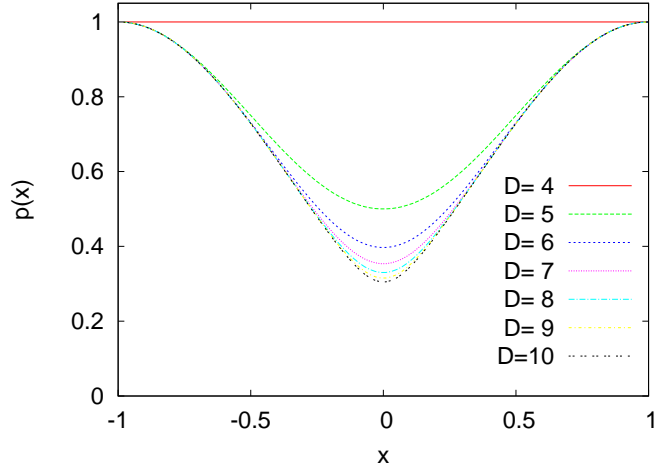}
}
\end{center}
\caption{The metric function $p$ is shown as a function of the  compactified coordinate $x$
for $4 \le D \le 10$ dimensions.
\label{Fig6}
}
\end{figure}

Solutions of the ODEs can be found for all $D\geq 4$,
\begin{eqnarray}
p & = & \frac{\cos^2 z}{4}
\left[ \left(\frac{1+\sin z}{\cos z}\right)^{D-3} + 
       \left(\frac{\cos z}{1+\sin z}\right)^{D-3} \right]^{\frac{2}{D-3}} \ ,
\label{solpd}\\
a & = &
-\frac{b_0}{(D-3)r_0^{D-3}}\left(\frac{\pi}{2}
-\arctan\left[\left(\frac{1+\sin z}{\cos z}\right)^{D-3}\right] \right)\ ,
\label{sola2} \\
\phi & = &
-\frac{Q}{(D-3)r_0^{D-3}}\left(\frac{\pi}{2}
-\arctan\left[\left(\frac{1+\sin z}{\cos z}\right)^{D-3}\right] \right)\ ,
\label{solphi} 
\end{eqnarray}
where $z=\arctan(l/r_0)$. 
The integration constants have been chosen such that $p$ is an even function
of $z$,
and  $a \to 0$ and $\phi\to 0$ as $l\to\infty$.
The parameter $b_0$ is related to the mass. We exhibit the function $p(x)$ for $4 \le D \le 10$
dimensions in Fig.~\ref{Fig6}.

\section{Asymptotic solutions and the slowly rotating limit}

\subsection{The $l \to 0$ expansion}

Restricting to symmetric wormhole solutions, the following expansion holds near the throat $l=0$ 
\begin{eqnarray}
a(l)=\sum_{k\geq 0} {  a_{2k}}{l^{2k}},~~q(l)=\sum_{k\geq 0} { q_{2k}}{l^{2k}}, 
\end{eqnarray}
with the first order coefficients
\begin{eqnarray}
&&
a_2=2c_\omega^2\frac{e^{4a_0-q_0}}{r_0^8},~~
q_2=\frac{2}{r_0}(e^{2(q_0-a_0)}-1),~~
a_4=\frac{c_\omega^2e^{2(q_0-a_0)}}{3r_0^{10}}
\left(\frac{8c_\omega^2e^{6a_0-q_0}}{r_0^6}
-2(5e^{2a_0}+e^{2q_0})
\right )
,
\\
\nonumber
&&
q_4=\frac{ e^{-4a_0}}{3r_0^4}
\left (
2(3  e^{4a_0}+2e^{4q_0}-5e^{2(a_0+q_0)} )
-\frac{4c_\omega^2e^{6a_0+q_0}}{r_0^6}
\right)~,
\end{eqnarray}
which contain two arbitrary parameters
$a_0$ and $q_0$. From the constraint equation (\ref{ric22}), 
it follows that these coefficients (together with the charge $Q$) fix the angular momentum
of the solutions,
\begin{eqnarray}
c_\omega=e^{\frac{q_0}{2}-2a_0}r_0^3 \sqrt{\frac{1}{2}(4-e^{-2a_0+2q_0})-\frac{8\pi G Q^2}{r_0^4}} \ .
\end{eqnarray}
The near-throat expression of the metric function $\omega(l)$ is
\begin{eqnarray}
\omega(l)=
\frac{4 e^{4a_0-q_0}c_\omega}{r_0^5} l
+\frac{2e^{4a_0-q_0}c_\omega}{3 r_0^7}(-9+8a_2r_0^2-2q_2r_0^2) l^3
+O(l^5).
\end{eqnarray}

\subsection{The $l \to \pm \infty$ expansion}
Considering again the case of  symmetric wormholes,
one finds the following asymptotic form of the solutions ($l \to \pm \infty$), 
\begin{eqnarray}
a(l)=\sum_{k\geq 1}\frac{\bar a_{2k}}{l^{2k}},~~q(l)=\sum_{k\geq 1}\frac{\bar q_{2k}}{l^{2k}}, 
\end{eqnarray}
involving two  parameters
$\bar q_2$ and $\bar q_4$.
The first coefficients in the above expansion are given by
\begin{eqnarray}
\bar a_2=\bar q_2,~~
\bar a_4=-\frac{\bar q_2r_0^2}{2},~~
\bar a_6=\frac{1}{48}(c_\omega^2+14 \bar q_2r_0^2),~~
\bar q_6=-\bar q_4 r_0^2-\frac{1}{96}c_\omega^2-\frac{5}{24}\bar q_2r_0^4 \ .
\end{eqnarray}
Note that, from (\ref{ric22}) the parameter $\bar q_4$ is fixed by $Q$ and $\bar q_2$,
\begin{eqnarray}
\bar q_4=-\frac{2}{3}\pi G Q^2+\frac{1}{8}(r_0^2-2\bar q_2)^2 \ .
\end{eqnarray}
For completeness, we give also the leading order terms in the asymptotic expansion
of the function $\omega(l)$ associated with rotation  
\begin{eqnarray}
\omega(l)=-\frac{c_\omega}{4 }\frac{1}{l^4}+\frac{c_\omega}{2}(\frac{r_0^2}{2}-q_2)\frac{1}{l^6}+O(1/l^8).
\end{eqnarray}

\subsection{The slowly rotating limit}

We suspect that the spinning wormhole solutions
could be constructed in closed form. 
However, despite our efforts, we
could not find  so far a closed form solution of the system (\ref{eqa})-(\ref{eqq}).
(An analytical solution is reported in the next Appendix;
however, that configuration has rather special properties.)

Nevertheless, some progress in this direction can be achieved in the slowly rotating limit.
Such solutions can be found by considering perturbation theory around the static  wormholes
in terms of the small parameter $c_\omega$. 
The second parameter of the solution is the throat radius $r_0$.

In this approach, the functions $a$ and $q$ have the following general expression
\begin{eqnarray}
\label{aq1}
a(l)=\sum_{k\geq 0}a_{2k}(l)c_\omega^{2k},~~q(l)=\sum_{k\geq 0}q_{2k}(l)c_\omega^{2k}.
\end{eqnarray}
Restricting again to symmetric wormholes, we give here the
expression for the first terms in this expansion
\begin{eqnarray}
a_0(l)=q_0(l)=0,~~
a_2(l)=-\frac{1}{r_0^4(2l^2+r_0^2)},~~
q_2(l)=\frac{1}{\pi r_0^6}
\left[
4-2\pi(1+\frac{2l^2}{r_0^2})
+\frac{16 l}{r_0^2}\sqrt{l^2+r_0^2}{\arctan}\frac{l}{l^2+r_0^2}
\right]~.
\end{eqnarray}
The leading order behaviour of the metric function associated with rotation is
\begin{eqnarray}
\omega(l)=\frac{2c_\omega}{r_0^4}\left(-1+\frac{2l \sqrt{l^2+r_0^2}}{2l^2+r_0^2} \right).
\end{eqnarray}
The higher order terms in (\ref{aq1}) have more complicated expressions, 
where we could not identify a general pattern.

\section{A  rotating wormhole exact solution}
\setcounter{equation}{0}

By using an `educated guess' approach,  we have found the following exact solution of the
  equations (\ref{eqa})-(\ref{eqq}):
\begin{eqnarray}
a(l)=\log \big( \frac{\sqrt{2}\sqrt{2l^2+r_0^2}}{\sqrt{ l^2+r_0^2}+l} \big)+a_0,~~~
q(l)=-2\log \big(\frac{l+\sqrt{ l^2+r_0^2}}{2^{1/4}r_0}\big)+a_0,~~~
\end{eqnarray}
with $a_0$ an arbitrary parameter and 
\begin{eqnarray}
c_\omega=\frac{e^{-\frac{3a_0}{2}}r_0^3}{2^{5/4}},~~Q=\frac{1}{4\sqrt{G}}\sqrt{\frac{3}{\pi}}r_0^2~.
\end{eqnarray}
{From} (\ref{omprime}) this implies
\begin{eqnarray}
\omega(l)=\frac{e^{\frac{3a_0}{2}}2^{5/4}r_0}{(l+\sqrt{ l^2+r_0^2})^2}+w_0.
\end{eqnarray}

This solution takes
a relatively simple form 
in terms of the metric parametrization (\ref{rot1})
\footnote{Note, that the equations (\ref{eqa})-(\ref{eqq})
have the following symmetries: 
$
i)~a\to a+a_0,~q\to q+a_0,
c_\omega \to c_\omega e^{-3 a_0}
$,
$ii)~r_0 \to  U r_0$
and
$iii)~l\to \lambda l$, 
$r_0 \to  \lambda r_0$,
$c_\omega \to c_\omega \lambda^3$
(with arbitrary nonzero parameters $U,\lambda$).
} 
(with  $a_0=w_0=0$ and $r_0=R_0^2 2^{1/4}$):
\begin{eqnarray}
\label{new-sol1}
&&
ds^2=-\frac{2(2r^2+R_0^2)}{(r+\sqrt{r^2+R_0^2})^2}dt^2
+\frac{(2r^2+R_0^2)(r+\sqrt{r^2+R_0^2})^2}{2(r^2+R_0^2)}
\left (dr^2+(r^2+R_0^2)\frac{1}{4}d\Omega_2^2 \right)
\\
\nonumber
&&
{~~~~~~~~~~~~~~~~~~~~~~~~~~~~~~~~~~~~~~~~~~}
+\frac{R_0^4}{8}
\left(
d\bar \psi+\cos \bar \theta d\bar \varphi-\frac{4}{(r+\sqrt{r^2+R_0^2})^2}dt
\right)^2,
\end{eqnarray} 
where we have introduced the scaled radial coordinate
  $r=l/(2^{1/4} R_0)$, 
  while
$d\Omega_2^2= d\bar \theta^2+\sin^2\bar \theta d \bar \varphi^2$.
The corresponding expression of the scalar field is
\begin{eqnarray}
\label{new-sol-scalar}
\phi=\frac{1}{2\sqrt{G}}\sqrt{\frac{3}{\pi}}
\bigg(
\arctan\frac{r}{\sqrt{r^2+R_0^2}}
-\frac{\pi}{4}
\bigg).
\end{eqnarray} 
It is clear that (\ref{new-sol1}) 
describes a rotating wormhole solution, the throat being located at $r=0$.
(Note the absence of a static limit.) 
A straightforward computation shows that both the Ricci and Kretschmann
scalars are finite for any value of $r$.
Also, a surface of constant $r,t$ has a spherical topology.

However, the asymptotics of this solutions are rather different as compared to the 
numerical solutions in this work.
For $r\to \infty$, we define a new radial coordinate $r=\sqrt{R}$, 
such that the asymptotic form of (\ref{new-sol1})
reads 
\begin{eqnarray}
\label{new-sol2}
ds^2=-dt^2+ dR^2+R^2d\Omega_2^2+\frac{R_0^4}{8}
(d\bar \psi+\cos \bar \theta d\bar \varphi )^2.
\end{eqnarray}
A three-dimensional
surface $R = const.,~t = const.$ has the topology of the Hopf bundle, 
$S^1$ fiber over $S^2$ base space.
Moreover, to leading order, (\ref{new-sol2})
shares the same asymptotics with a Gross-Perry-Sorkin Kaluza-Klein monopole \cite{Gross:1983hb}.
Then, for $r>0$, the solution (\ref{new-sol1}), (\ref{new-sol-scalar}) can be considered 
as the wormhole counterpart of the rotating Kaluza-Klein black holes
with squashed horizon discussed $e.g.$ in \cite{Nakagawa:2008rm}.

Even more complicated asymptotics are found for $r \to -\infty$.
Defining $r=-\sqrt{R}$, one finds as $R\to \infty$ the line element
\begin{eqnarray}
\label{new-sol3}
ds^2=-\frac{16}{R_0^4} R^2dt^2+ \frac{R_0^4 }{16} \frac{ dR^2 }{ R^2}+\frac{R_0^4}{16}d\Omega_2^2+\frac{R_0^4}{8}
(d\bar \psi+\cos \bar \theta d\bar \varphi -\frac{16}{R_0^4}Rdt)^2.
\end{eqnarray}
This geometry
describes a fibration of $AdS_2$ over the homogeneously squashed $S^3$ with symmetry
group $SO(2, 1) \times SU(2) \times U(1)$.

Thus it is natural to interpret the solution (\ref{new-sol1}), (\ref{new-sol-scalar}) 
as describing a  Lorentzian wormhole interpolating between a 
Kaluza-Klein monopole background and a squashed
 $AdS_2\times S^3$ spacetime.

\end{appendix}

%
\end{document}